\documentclass[aps,prd,nofootinbib,showkeys,preprint,floatfix]{revtex4}

\usepackage{amsmath}
\usepackage{amssymb}
\usepackage{graphicx}
\usepackage{rotating}
\usepackage{color} %This is essential for placing graphics onto
\usepackage{multirow} % To use multirow command in tables
 \usepackage[T1]{fontenc}

\newcommand{\nn}{\nonumber}

\newcommand{\AddrAHEP}{
  {\it AHEP Group, Instituto de F\'{\i}sica Corpuscular --
    C.S.I.C./Universitat de Val{\`e}ncia \\
    Edificio de Institutos de Paterna, Apartado 22085,
  E--46071 Val{\`e}ncia, Spain}}

\newcommand{\AddrLisb}{%
 Departamento de F\'\i sica and CFTP, Instituto Superior T\'ecnico\\
          Av. Rovisco Pais 1, 1049-001 Lisboa, Portugal }

\newcommand{\AddrWur}{%
Institut f\"ur Theoretische Physik und Astronomie, 
Universit\"at W\"urzburg\\
Am Hubland, 
97074 Wuerzburg}

\def\gsim{\raise0.3ex\hbox{$\;>$\kern-0.75em\raise-1.1ex\hbox{$\sim\;$}}}
\def\lsim{\raise0.3ex\hbox{$\;<$\kern-0.75em\raise-1.1ex\hbox{$\sim\;$}}}

\begin{document}

\preprint{CFTP/10-013}  
\preprint{IFIC/10-42}  

\title{Supersymmetric type-III seesaw: \\ lepton flavour 
violating decays and dark matter}

\author{J. N. Esteves}\email{joaomest@cftp.ist.utl.pt}\affiliation{\AddrLisb}
\author{M.~Hirsch} \email{mahirsch@ific.uv.es}\affiliation{\AddrAHEP}

\author{F. Staub}\email{florian.staub@physik.uni-wuerzburg.de}
\affiliation{\AddrWur}

\author{W. Porod} \email{porod@physik.uni-wuerzburg.de}\affiliation{\AddrWur}
\author{J.~C.~Romao}\email{jorge.romao@ist.utl.pt}\affiliation{\AddrLisb}



\begin{abstract}
We study a supersymmetric version of the seesaw mechanism type-III. 
The model consists of the MSSM particle content plus three copies of 
${\bf 24}$ superfields. The fermionic part of the $SU(2)$ triplet 
contained in the ${\bf 24}$ is responsible for the type-III seesaw, 
which is used to explain the observed neutrino masses and mixings. 
Complete copies of ${\bf 24}$ are introduced to maintain gauge coupling 
unification. These additional states change the beta functions of the gauge
couplings above the seesaw scale.
Using mSUGRA boundary conditions we calculate the 
resulting supersymmetric mass spectra at the electro-weak scale using 
full 2-loop renormalization group equations. We show that the resulting
spectrum can be quite different compared to the usual mSUGRA spectrum.
We discuss how this might be used to obtain information
on the seesaw scale from mass measurements.
 Constraints on the 
model space due to limits on lepton flavour violating decays are 
discussed. The main constraints come from the bounds on $\mu\to e \gamma$
but there are also regions where the decay $\tau\to \mu \gamma$ gives
stronger constraints.
We also calculate the regions allowed by the dark 
matter constraint. For the sake of completeness, we compare our results 
with those for the supersymmetric seesaw type-II and, to some extent, 
with type-I. 
\end{abstract}

\maketitle

%\tableofcontents

\section{Introduction}
\label{sec:int}

Supersymmetry offers a number of advantages compared to the standard
model (SM). To name just a few, SUSY has a dark matter candidate, it
can alleviate the gauge hierarchy problem and the minimal
supersymmetric extension of the standard model (MSSM) leads to gauge
coupling unification, if SUSY particles exist with masses of the order
of the electro-weak scale. However, in the MSSM neutrino masses are
zero, just as in the SM. Neutrino oscillation experiments
\cite{Fukuda:1998mi,Ahmad:2002jz,Eguchi:2002dm,KamLAND2007}, on the
other hand, have shown that neutrinos have tiny, but non-zero, masses
and that mixing in the leptonic sector is large \cite{Schwetz:2008er}.

From a theoretical point of view, if neutrinos are Majorana particles, 
{\em all} models of neutrino mass at low energies reduce to the unique 
dimension-5 operator \cite{Weinberg:1979sa}
\begin{equation}\label{eq:dim5}
(m^{\nu})_{\alpha\beta} = \frac{f_{\alpha\beta}}{\Lambda} (H L) (H L) \thickspace .
\end{equation}
Neutrino experiments determine only $f_{\alpha\beta}/\Lambda$, but 
contain no information about the origin of this operator, nor about 
the absolute size of $\Lambda$. If $f$ is a coefficient ${\cal O}(1)$, 
current neutrino data indicates $\Lambda \lsim {\cal O}(10^{15})$ GeV. 
This is the essence of the ``seesaw'' mechanism. 

One can show that there are exactly three different tree-level
realizations of the seesaw mechanism
\cite{Ma:1998dn}. Type-I is the well-known case of the exchange of 
a heavy fermionic singlet \cite{Minkowski:1977sc,seesaw,MohSen}. 
Type-II corresponds to the exchange of a scalar $SU(2)$ triplet 
\cite{Schechter:1980gr,Cheng:1980qt}. In seesaw type-III 
one adds (at least two) fermionic $SU(2)$ triplets to the field
content of the SM \cite{Foot:1988aq}.  If in case of type-II and 
type-III models one would extend the usual MSSM by just the superfields
responsible for neutrino masses and mixings, one would destroy the
nice feature of gauge coupling unification as they belong to
incomplete $SU(5)$ representations. This problem is easily cured by
embedding the new states in complete $SU(5)$ representations, e.g.~in
case of type-II in {\bf 15}-plets \cite{Rossi:2002zb} and in case of type
III in {\bf 24}-plets \cite{Buckley:2006nv}. Note, that the {\bf 24}-plet contains
beside the $SU(2)$ triplet also a singlet state which also contributes
to neutrino physics and, thus, one has in this case actually a mixture
between type-I and type-III.

Understanding the nature of supersymmetry breaking by measuring the
soft parameters will be one of the central tasks if signals of SUSY
are found at the LHC.  All the more so, since one can possibly gain
some insight into the high energy scale physics from such
measurements. Two kind of measurements containing indirect information
about the seesaw scale in SUSY models exist in principle: lepton
flavour violating (LFV) observables and sparticle masses. In case of
seesaw type-I, low energy LFV decays such as $l_i \to l_j +\gamma$ and
$l_i \to 3 l_j$ have been calculated in
\cite{Hisano:1995nq,Hisano:1995cp,Ellis:2002fe,Deppisch:2002vz,Petcov:2003zb,
Arganda:2005ji,Petcov:2005yh,Antusch:2006vw,Deppisch:2004fa,Hirsch:2008dy}; 
$\mu-e$ conversion in nuclei has been studied in 
\cite{Arganda:2007jw,Deppisch:2005zm}. The type-II model has 
received less attention, although it has actually fewer free
parameters than type-I implying that ratios of LFV decays of leptons
can actually be predicted as a function of neutrino angles in mSUGRA,
as has been shown in \cite{Rossi:2002zb,Hirsch:2008gh}. 
A first study has been done in \cite{Biggio:2010me}. We stress that such a 
setup can not explain neutrino data unless non-renormalizable operators are added, as 
indeed is done in \cite{Biggio:2010me}. This is due to the need of generating a sufficiently 
large splitting between the Yukawa couplings of the singlet and the 
triplets, which can not be obtained from RGE running only.  Moreover, 
in the above publications for the type-II and type-III models only 1-loop 
RGEs have been used. However, we will show that using 2-loop RGEs 
is important for the calculation of the spectrum as this leads to a 
shift of the GUT scale.

Measurements at colliders, once SUSY is discovered, can provide
additional information. LFV decays of left sleptons within mSUGRA have
been studied for type-I in \cite{Hisano:1998wn,Esteves:2009vg} and for
type-II in \cite{Hirsch:2008gh,Esteves:2009vg}. Precise mass
measurements, in particular of the sleptons and sneutrinos, might also
show indirect effects of the seesaw
\cite{Blair:2002pg,Freitas:2005et,Deppisch:2007xu}. As mentioned
above, the additional heavy states of type-II and type-III lead to
changes in the running of the beta functions and also of the mass
parameters above the seesaw scale leading to changes of the spectrum
at the electro-weak scale compared to the usual mSUGRA expectations.
From different combinations of masses one can form ``invariants'',
i.e. numbers which to leading order depend only on the seesaw scale
\cite{Buckley:2006nv}, although there are important corrections at
2-loop for the type-II \cite{Hirsch:2008gh} and, as we will show in
this paper, also for type-III. It is also interesting to note, that
the additional Yukawa couplings at the high scale can lead to a mass
splitting between smuons and selectrons which in principle can be
measured at the LHC: it has been shown in ref.~\cite{Allanach:2008ib}
that such a splitting may be constrained down to ${\cal O}(10^{-4})$
for 30 $fb^{-1}$ of integrated luminosity. In mSUGRA, one expects this
splitting to be tiny, whereas in mSUGRA plus seesaw significantly
different masses are generated, as has been shown for type-I in
ref.~\cite{Abada:2010kj}.

The modified spectrum also affects the calculation of the relic
density.  Assuming the standard thermal history of the early universe
only four very specific regions in parameter space of mSUGRA can
correctly explain the most recent WMAP data
\cite{Komatsu:2010fb}. These are (i) the bulk region; (ii) the
co-annihilation line; (iii) the ``focus point'' line and (iv) the
``Higgs funnel'' region. In the bulk, where the SUSY particles are
relatively light, no specific relations among the sparticle masses
exist. In the co-annihilation line the lightest scalar tau is nearly
degenerate with the lightest neutralino, thus reducing the neutralino
relic density with respect to naive expectations
\cite{Griest:1990kh,Baer:2002fv}. In the ``focus point'' line
\cite{Feng:1999zg,Baer:2002fv} $\Omega_{\tilde \chi^0_1}h^2$ is small enough
to explain $\Omega_{DM}h^2$ due to a rather small value of $\mu$
leading to an enhanced higgsino component in the lightest neutralino
and thus an enhanced coupling to the $Z$-boson. Lastly, at large
$\tan\beta$ an s-channel resonance pair annihilation of neutralinos
through the CP-odd Higgs boson can become important. This is called
the ``Higgs funnel'' region \cite{Drees:1992am}.  Also in the seesaw
models of type-II and III these regions exist but the regions get
shifted. Moreover, if the seesaw scale is sufficiently low the
co-annihilation region disappears in type-II models
\cite{Esteves:2009qr}.  We will show that the same happens in case of
the type-III model and we will contrast the results of this model with
type-I and type-II models.

The rest of this paper is organized as follows. In the next section, we
first define the model. For completeness, and since
we will compare the results for the different variants, we give the
definitions for minimal type-I and type-II seesaws as well. We have
used SARAH \cite{Staub:2008uz,Staub:2009bi,Staub:2010jh} to calculate
the full 2-loop RGEs, based on the general expressions given in
\cite{Martin:1993zk}. We have, where possible, compared our results to
previously available work and generally found agreement. However,
\cite{Borzumati:2009hu} have calculated 1-loop RGEs for all parameters
and found some differences in case of the seesaw type-II to the RGEs
published in \cite{Rossi:2002zb}. Our calculation agrees with
\cite{Borzumati:2009hu}. We then turn to the discussion of the
resulting SUSY spectrum. The large changes in the spectrum affects the
predictions for the rates of rare lepton decays, such as $\mu\to e
\gamma$, and the relic density as discussed in section
\ref{sec:res}. We present in section
\ref{sec:cncl} our conclusions. In the appendix we first summarize
the procedure on how to obtain the RGEs for the soft SUSY breaking
parameters from the beta functions and anomalous dimensions.
We then give the formulas at the 1-loop and 2-loop level for
these quantities for the seesaw models of type-II and type-III for
an arbitrary number of new seesaw particles which are decomposed according
to their SM gauge quantum numbers.

\section{Models and spectra}
\label{sec:model}

In this section we briefly recall the main features of the three
seesaw models.  In models of type-II and III one adds particles
charged under the SM gauge group.  As they correspond to incomplete
$SU(5)$ representations, they would destroy the nice feature of gauge
coupling unification. For this reason we add at the seesaw scale(s)
additional particles to obtain complete $SU(5)$ representations which
we briefly review below.  A more detailed discussion including the
embedding in $SU(5)$ models can be found in \cite{Borzumati:2009hu}.

In the subsequent sections we present the various superpotentials. In
addition there will also be the corresponding soft SUSY terms which,
however, reduce at the electro-weak scale to the MSSM one and, thus,
are not discussed further. The additional terms of the soft SUSY
breaking potential, due to the heavy particles, do not effect the
discussion presented later on, as their effect is at most of the order
$M_{EWSB}/M_{seesaw}$ and, thus, can be safely neglected. In this
paper we will assume common soft SUSY breaking at the GUT-scale
$M_{GUT}$ to specify the spectrum at the electro-weak scale: a common
gaugino mass $M_{1/2}$, a common scalar mass $m_0$ and the trilinear
coupling $A_0$ which gets multiplied by the corresponding Yukawa
couplings to obtain the trilinear couplings in the soft SUSY breaking
Lagrangian. In addition the sign of the $\mu$ parameter is fixed, as
is $\tan\beta =v_u/v_d$ at the electro-weak scale, where $v_d$ and
$v_u$ are the the vacuum expectation values (vevs) of the neutral
component of $H_d$ and $H_u$, respectively. The models discussed below
also contain new bilinear parameters in the superpotential leading to
additional bilinear terms in the soft SUSY breaking potential which
are proportional to $B_0$ of the MSSM Higgs sector. The corresponding
RGEs decouple and their only effect is a small mass splitting between
the new heavy scalar particles from the new heavy fermionic states of
the order $B_0/M_{seesaw}$. This leads to a tiny effect in the
calculation of the thresholds at the seesaw scale(s)
\cite{Kang:2010zd} which, however, we can safely neglect.

\subsection{Supersymmetric seesaw type-I}
\label{sec:modelI}

In the case of seesaw type-I one postulates very heavy right-handed neutrinos
yielding the following superpotential below $M_{GUT}$:
\begin{eqnarray}
W_{I} &=& W_{MSSM} + W_{\nu} \label{eq:superpotI} \thickspace, \\
W_{MSSM}& = & {\widehat U}^c Y_u {\widehat Q} \cdot {\widehat H}_u
         - {\widehat D}^c Y_d {\widehat Q} \cdot {\widehat H}_d
         - {\widehat E}^c Y_e {\widehat L} \cdot {\widehat H}_d 
              + \mu {\widehat H}_u \cdot {\widehat H}_d \thickspace, \\ 
 W_{\nu}& = &  {\widehat N}^c Y_\nu {\widehat L} \cdot {\widehat H}_u
          + \frac{1}{2}{\widehat N}^c M_R  {\widehat N}^c \thickspace,
\end{eqnarray}
where $A \cdot B = A_1 B_2 - A_2 B_1$ denotes the $SU(2)$ invariant product of 
two $SU(2)$ doublets. This model can be embedded in an $SU(5)$ using
the following $SU(5)$ matter representations: $ 1= N^c$, ${\bar 5}_M=\{D^c,L\}$
and $10_M=\{Q,U^c,E^c\}$. 
For the neutrino mass matrix one obtains the well-known formula
\begin{equation}
m_\nu = - \frac{v^2_u}{2} Y^T_\nu M^{-1}_R Y_\nu.
\label{eq:mnuI}
\end{equation}
Being complex symmetric, the light Majorana neutrino mass matrix
  in eq.~(\ref{eq:mnuI}), is diagonalized by a unitary $3\times 3$ matrix
  $U$~\cite{Schechter:1980gr}
\begin{equation}\label{diagmeff}
{\hat m_{\nu}} = U^T \cdot m_{\nu} \cdot U\ .
\end{equation}

Inverting the seesaw equation, eq.~(\ref{eq:mnuI}), allows to express 
$Y_{\nu}$ as \cite{Casas:2001sr}
\begin{equation}\label{Ynu}
Y_{\nu} =\sqrt{2}\frac{i}{v_u}\sqrt{\hat M_R}\cdot R \cdot \sqrt{{\hat
    m_{\nu}}} \cdot U^{\dagger},
\end{equation}
where the $\hat m_{\nu}$ and $\hat M_R$ are  diagonal matrices containing the
corresponding eigenvalues. 
$R$ is  in general a complex orthogonal matrix. Note that, in the
special case $R={\bf 1}$, $Y_{\nu}$ contains only ``diagonal'' products
$\sqrt{M_im_{i}}$. For $U$ we will use the standard form
\begin{eqnarray}\label{def:unu}
U=
\left(
\begin{array}{ccc}
 c_{12}c_{13} & s_{12}c_{13}  & s_{13}e^{-i\delta}  \\
-s_{12}c_{23}-c_{12}s_{23}s_{13}e^{i\delta}  & 
c_{12}c_{23}-s_{12}s_{23}s_{13}e^{i\delta}  & s_{23}c_{13}  \\
s_{12}s_{23}-c_{12}c_{23}s_{13}e^{i\delta}  & 
-c_{12}s_{23}-s_{12}c_{23}s_{13}e^{i\delta}  & c_{23}c_{13}  
\end{array}
\right) 
 \times
 \left(
 \begin{array}{ccc}
 e^{i\alpha_1/2} & 0 & 0 \\
 0 & e^{i\alpha_2/2}  & 0 \\
 0 & 0 & 1
 \end{array}
 \right)
\end{eqnarray}with $c_{ij} = \cos \theta_{ij}$ and $s_{ij} = \sin \theta_{ij}$. The angles $\theta_{12}$,
 $\theta_{13}$ and  $\theta_{23}$ are the solar neutrino angle, the reactor (or CHOOZ) angle
and the atmospheric neutrino mixing angle, respectively. $\delta$ is the Dirac phase
and $\alpha_i$ are Majorana phases. In the following we will set the latter to 0 and
consider in case of $\delta$ only the cases $0$ and $\pi$.

\subsection{Supersymmetric seesaw type-II}
\label{sec:modelII}

In seesaw models of type-II one adds a scalar $SU(2)$ triplet $T$ to generate
neutrino masses. As this triplet carries also hypercharge one has to embed
it in a $15$-plet of $SU(5)$ which has under
 $SU(3)\times SU(2) \times U(1)$ the following decomposition \cite{Rossi:2002zb}
\begin{eqnarray}\label{eq:15}
{\bf 15} & = &  S + T + Z  \thickspace,\\ \nonumber
S & \sim  & (6,1,-\frac{2}{3}), \hskip10mm
T \sim (1,3,1), \hskip10mm
Z \sim (3,2,\frac{1}{6}).
\end{eqnarray}
One has to add two {\bf 15}-plets $15$ and $\overline{15}$ to avoid a chiral
anomaly below the GUT-scale.
The $SU(5)$ invariant superpotential reads as 
\begin{eqnarray}\label{eq:pot15}
W & = & \frac{1}{\sqrt{2}}{\bf Y}_{15} {\bar 5} \cdot 15 \cdot {\bar 5} 
   + \frac{1}{\sqrt{2}}\lambda_1 {\bar 5}_H \cdot 15 \cdot {\bar 5}_H 
+ \frac{1}{\sqrt{2}}\lambda_2 5_H \cdot \overline{15} \cdot 5_H 
+ {\bf Y}_5 10 \cdot {\bar 5} \cdot {\bar 5}_H \nonumber \\ 
 & + & {\bf Y}_{10} 10 \cdot 10 \cdot 5_H + M_{15} 15 \cdot \overline{15} 
+ M_5 {\bar 5}_H \cdot 5_H
\end{eqnarray}
with ${5}_H =(H^c,H_u)$ and 
${\bar 5}_H=({\bar H}^c,H_d)$. We do not show the part 
responsible for the $SU(5)$ breaking as we take the $SU(5)$ only 
as a guideline to fix
some of the boundary conditions at $M_{GUT}$.
Below $M_{GUT}$ in the $SU(5)$-broken 
phase the superpotential reads
\begin{eqnarray}\label{eq:broken}
W_{II} & = & W_{MSSM} + \frac{1}{\sqrt{2}}(Y_T \widehat{L} \widehat{T}_1  \widehat{L} 
+  Y_S \widehat{D}^c \widehat{S}_1 \widehat{D}^c) 
+ Y_Z \widehat{D}^c \widehat{Z}_1 \widehat{L}   \nonumber \\
& + & \frac{1}{\sqrt{2}}(\lambda_1 {\widehat H}_d \widehat{T}_1  {\widehat H}_d 
+\lambda_2 {\widehat H}_u \widehat{T}_2 {\widehat H}_u) 
+ M_T \widehat{T}_1 \widehat{T}_2 
+ M_Z \widehat{Z}_1 \widehat{Z}_2 + M_S \widehat{S}_1 \widehat{S}_2
\end{eqnarray}
where fields with index 1 (2) originate from the $15$-plet ($\overline{15}$-plet).
The second term in eq.~(\ref{eq:broken}) is responsible for the generation 
of the neutrino masses yielding 
\begin{eqnarray}\label{eq:ssII}
m_\nu=\frac{v_u^2}{2} \frac{\lambda_2}{M_T}Y_T.
\end{eqnarray}
Note that 
\begin{equation}\label{diagYT}
{\hat Y}_T = U^T \cdot Y_T \cdot U \thickspace,
\end{equation}
i.e. $Y_T$ is diagonalized by the same matrix as $m_{\nu}$. 
If all neutrino eigenvalues, angles and phases were known, $Y_T$ 
would be fixed up to an overall constant which can be easily 
estimated to be 
\begin{equation}\label{est}
\frac{M_T}{\lambda_2} \simeq 10^{15} {\rm GeV} \hskip2mm 
\Big(\frac{0.05 \hskip1mm {\rm eV}}{m_{\nu}}\Big).
\end{equation}

In addition there are the couplings $Y_S$ and $Y_Z$, which, in  
principle,  are not determined by any low-energy data. In the calculation 
of LFV observables in supersymmetry both matrices, $Y_T$ and 
$Y_Z$, contribute. Having a GUT model in mind we require for
the numerical discussion later the $SU(5)$ boundary conditions,
apart from threshold corrections, $Y_T=Y_S=Y_Z$ at $M_{GUT}$.  

As long as $M_Z \sim M_S \sim M_T \sim M_{15}$ gauge coupling unification 
will be maintained. The equality need not be exact for successful unification. 
In our numerical studies we have taken into account the different running
of these mass parameters but we decouple them all at the scale $M_T(M_T)$
because the differences are small.

\subsection{Supersymmetric seesaw type-III}
\label{sec:modelIII}

In the case of a seesaw model type-III one needs new fermions $\Sigma$ 
at the high
scale belonging to the adjoint representation of $SU(2)$. This has to be 
embedded in a {\bf 24}-plet  to obtain a complete $SU(5)$  representation.
The superpotential of the unbroken $SU(5)$ relevant for our discussion is
\begin{eqnarray}\label{eq:spot5}
W & = & \sqrt{2} \, {\bar 5}_M Y^5 10_M {\bar 5}_H 
          - \frac{1}{4} 10_M Y^{10} 10_M 5_H 
  +  5_H 24_M Y^{III}_N{\bar 5}_M +\frac{1}{2} 24_M M_{24}24_M \thickspace.
\end{eqnarray}
As above, we have not specified the Higgs sector responsible for the
$SU(5)$ breaking.
The new parts, which will give the seesaw mechanism, comes from 
the $24_M$. It decomposes under  $SU(3)\times SU(2) \times U(1)$ as 
\begin{eqnarray}\label{eq:def24}
24_M & = &(1,1,0) + (8,1,0) + (1,3,0) + (3,2,-5/6) + (3^*,2,5/6) \thickspace, \\ \nn
   & = & \widehat{B}_M + \widehat{G}_M + \widehat{W}_M + \widehat{X}_M + \widehat{\bar X}_M \thickspace.
\end{eqnarray}
The fermionic components of $(1,1,0)$ and $(1,3,0)$ have exactly 
the same quantum numbers as $\nu^c$ and $\Sigma$. Thus, the $24_M$ 
always produces a combination of the type-I and type-III seesaw. 

In the $SU(5)$ broken phase the superpotential becomes
\begin{eqnarray}\label{eq:spotIII}
 W_{III} & = &  W_{MSSM}
 +  \widehat{H}_u( \widehat{W}_M Y_N - \sqrt{\frac{3}{10}} 
               \widehat{B}_M Y_B) \widehat{L}
 + \widehat{H}_u \widehat{\bar X}_M Y_X \widehat{D}^c \nonumber \\
         & & + \frac{1}{2} \widehat{B}_M M_{B} \widehat{B}_M 
         + \frac{1}{2}\widehat{G}_M M_{G} \widehat{G}_M 
          + \frac{1}{2} \widehat{W}_M M_{W} \widehat{W}_M 
          + \widehat{X}_M M_{X} \widehat{\bar X}_M 
\end{eqnarray}
As before we use at the GUT scale the boundary condition
$Y_N = Y_B = Y_X$ and $M_B = M_G=M_W=M_X$. Integrating out the heavy fields
yields the following formula for the neutrino masses at the low scale:
\begin{equation}
m_\nu = - \frac{v^2_u}{2} 
\left( \frac{3}{10} Y^T_B M^{-1}_B Y_B + \frac{1}{2} Y^T_W M^{-1}_W Y_W \right). 
\label{eq:mnu_seesawIII}
\end{equation}
As mentioned above there are two contributions stemming from the gauge singlet
as well as from the $SU(2)$ triplet. In this case the calculation of the Yukawa
couplings in terms of a given high scale spectrum is more complicated than in the
other two types of seesaw models. However, as we start from universal couplings
and masses at  $M_{GUT}$ we find that at the seesaw scale one still has
$M_B \simeq M_W$ and $Y_B \simeq Y_W$ so that one can write in a good approximation
\begin{equation}
m_\nu = - v^2_u  \frac{4}{10} Y^T_W M^{-1}_W Y_W 
\label{eq:mnu_seesawIIIa}
\end{equation}
and one can use the corresponding  decomposition for $Y_W$ as discussed in section
\ref{sec:modelI} up to the overall factor $4/5$.

\subsection{Effects of the heavy particles on the MSSM spectrum}

\begin{figure}[t]
  \centering
  \begin{tabular}{cc}
\includegraphics[width=0.48\textwidth]{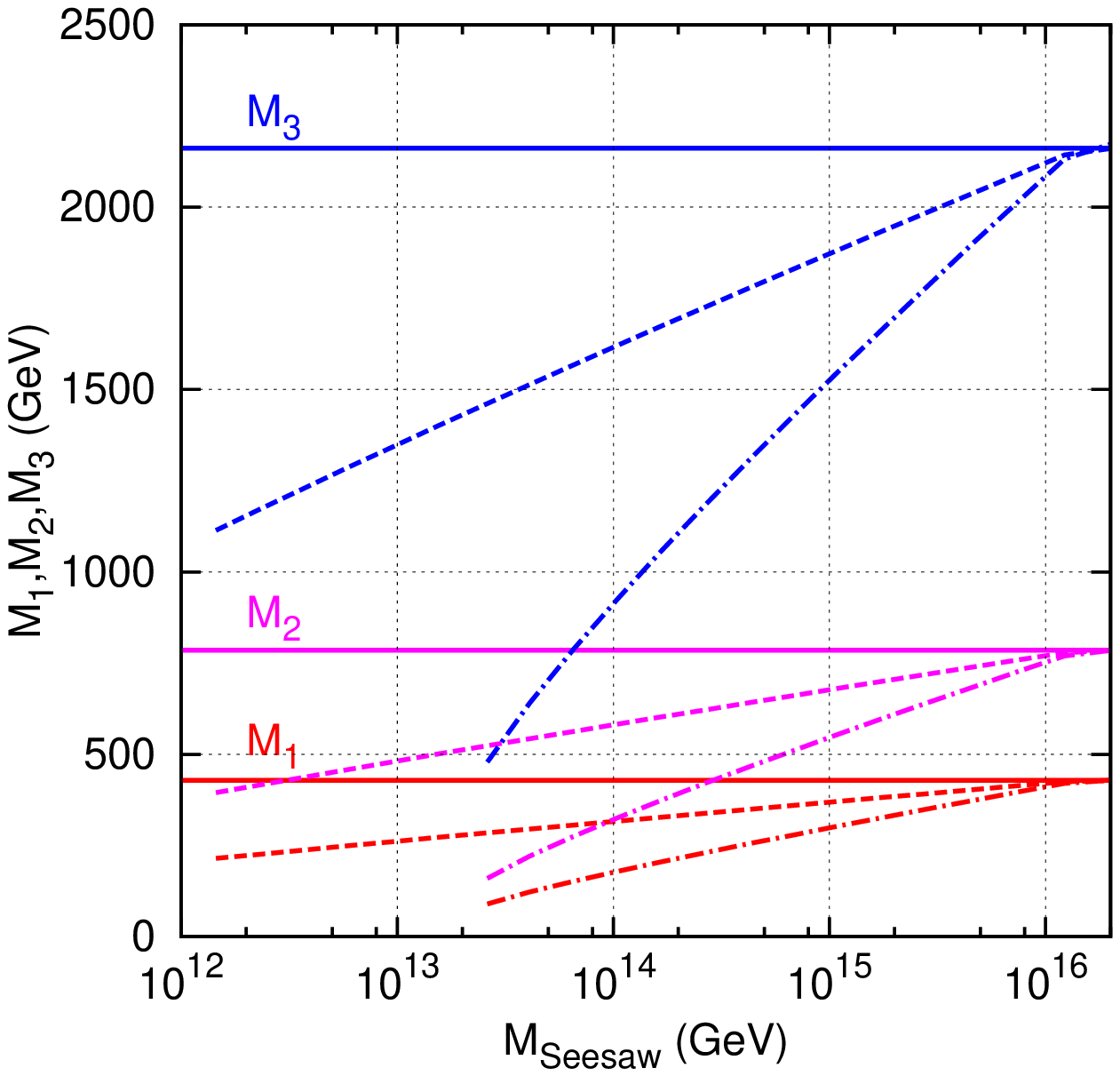}
\includegraphics[width=0.48\textwidth]{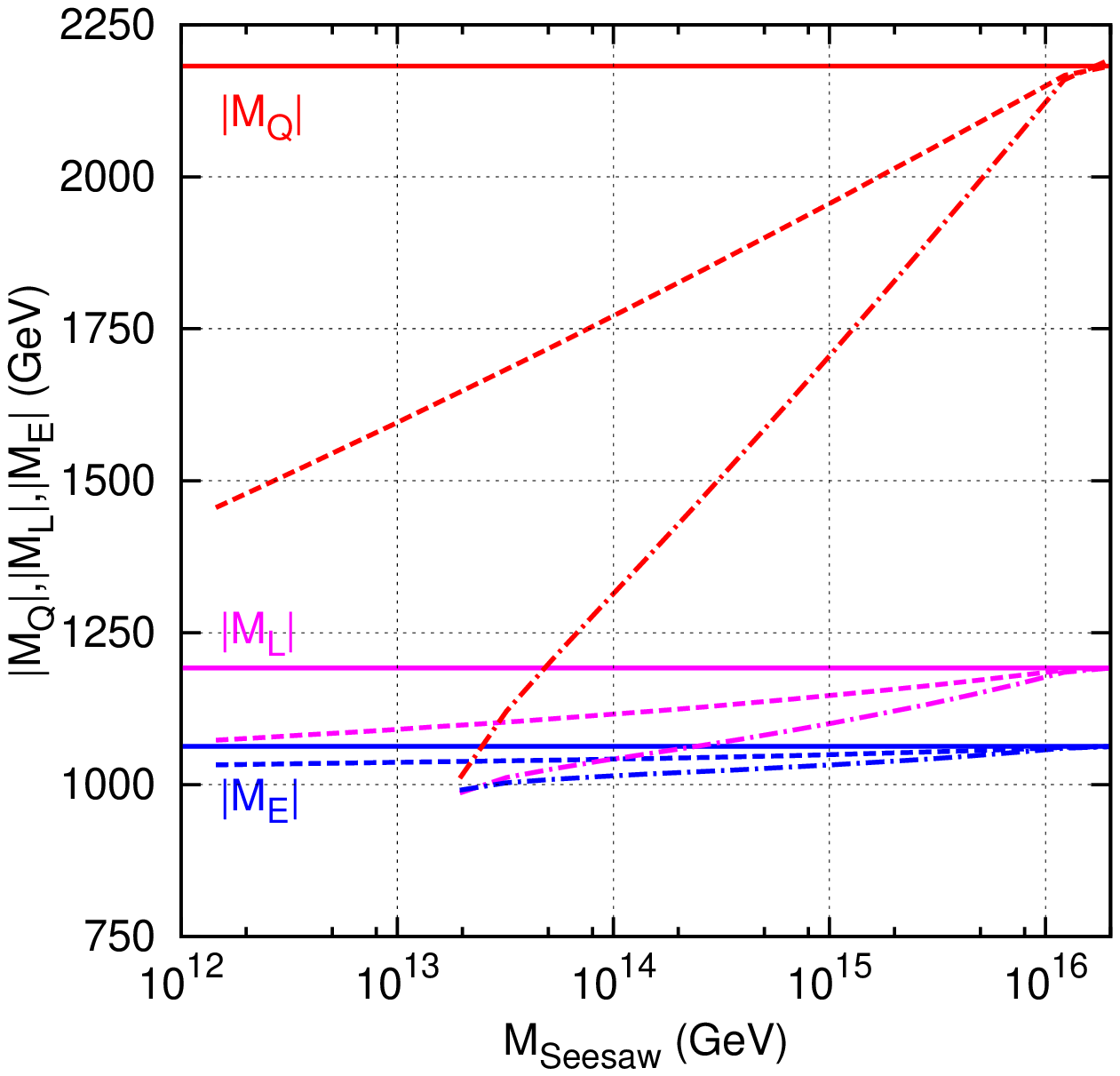}
\end{tabular}
\vspace{-5mm}
\caption{Mass parameters at $Q=1$ TeV versus the seesaw scale for fixed
high scale parameters $m_0= M_{1/2} = 1$ TeV, $A_0=0$, $\tan\beta=10$ and $\mu>0$. 
The full lines correspond to seesaw type-I, the dashed ones to type-II and the
dash-dotted ones to type-III. In all cases a degenerate spectrum of the seesaw
particles has been assumed.}
  \label{fig:mass1000}
\end{figure}

\begin{figure}[t]
  \centering
  \begin{tabular}{cc}
\includegraphics[width=0.48\textwidth]{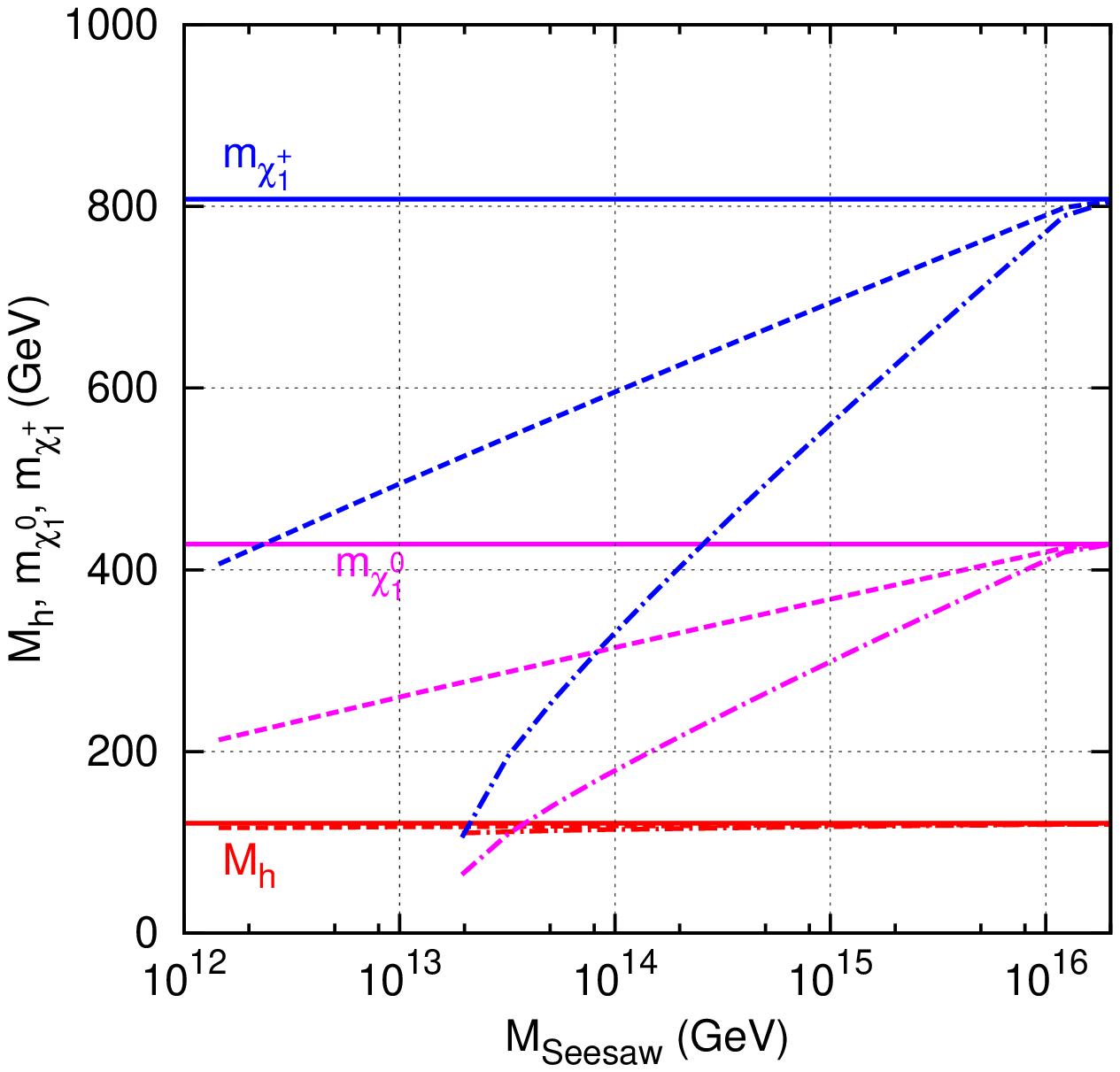}
\includegraphics[width=0.48\textwidth]{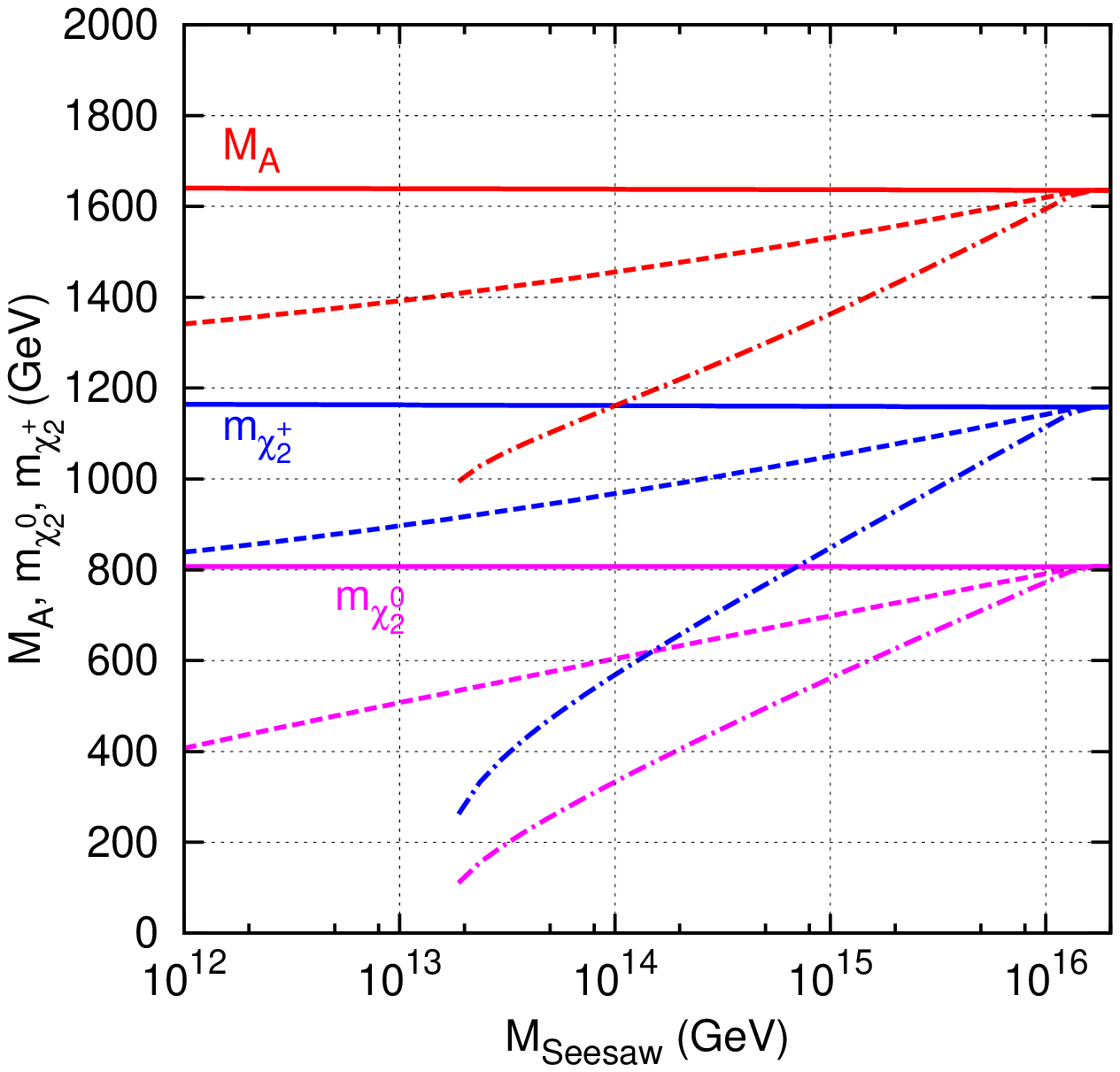}
\end{tabular}
\vspace{-5mm}
\caption{Example of spectra at $Q=1$ TeV versus the seesaw scale for fixed
high scale parameters $m_0= M_{1/2} = 1$ TeV, $\tan\beta=10$ and
$\mu>0$. On left panel $M_h, m_{\tilde \chi^0_1}, m_{\tilde \chi^+_1}$
while on the right panel we have  $M_A, m_{\chi^0_2},
m_{\tilde \chi^+_2}$. The line codes are as in Fig.~\ref{fig:mass1000}.}
  \label{fig:spectra}
\end{figure}

The appearance of charged particles at scales between the electro-weak
scale and the GUT scale leads to changes in the beta functions
of the gauge couplings \cite{Rossi:2002zb,Buckley:2006nv}.
In the MSSM the corresponding values at 1-loop level are
$(b_1,b_2,b_3)=(33/5,1,-3)$. In case of one {\bf 15}-plet the additional
contribution is $\Delta b_i=7/2$ whereas in case of {\bf 24}-plet
it is $\Delta b_i=5$. This results in case of type-II in 
a total shift of  $\Delta b_i=7$ for the minimal model and
in case of type-III in  $\Delta b_i=15$  assuming 3 generations
of {\bf 24}-plets. This does not only change the evolution
of the gauge couplings but also the evolution of the gaugino
and scalar mass parameters with profound implications on
the spectrum \cite{Buckley:2006nv,Hirsch:2008gh}. Additional
effects on the spectrum of the scalars can be present if some of the Yukawa 
couplings get large
\cite{Hirsch:2008gh,Calibbi:2009wk,Biggio:2010me}
which can also happen in type-I models \cite{Calibbi:2007bk}.
In Fig.\ \ref{fig:mass1000} we exemplify this by showing the values
of selected mass parameters at $Q=1$ TeV versus the seesaw scale for
fixed high scale parameters $m_0= M_{1/2} = 1$ TeV and we have set the
additional Yukawa couplings to zero. As expected, the effects in case
of models of type-II and III are larger the smaller the corresponding
seesaw-scale is. The scalar mass parameters shown are of the first 
generation and, thus, the results are nearly independent of $\tan\beta$
and $A_0$. For illustration we show in Fig.~\ref{fig:spectra} the corresponding
spectrum where we have fixed $\tan\beta=10$ and $A_0=0$.

\begin{figure}[t] \centering
\includegraphics[height=60mm,width=80mm]{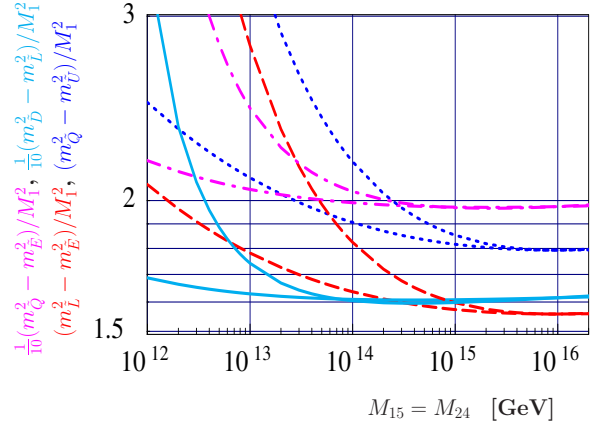}
\vskip0mm
\caption{Four different ``invariant'' combinations of soft masses 
versus the mass of the ${\bf 15}$-plet {\em or} ${\bf 24}$-plet, 
$M_{15}=M_{24}$. The plot assumes that the Yukawa couplings are 
negligibly small. The calculation is at 1-loop order in the leading-log 
approximation. The lines running faster up towards smaller $M$ are 
for type-III seesaw, the values for type-II seesaw are shown for 
comparison.}
\label{fig:ana}
\end{figure}

We note that in all three model types the ratio of the
gaugino mass parameters is nearly the same as in the usual mSUGRA scenarios
but the ratios  of the sfermion mass parameters change
\cite{Buckley:2006nv,Hirsch:2008gh}. One can form four 'invariants' for which at
least at the 1-loop level the dependence on $M_{1/2}$ and $m_0$ is rather weak,
e.g.\ $(m^2_L - m^2_E)/M^2_1$, $(m^2_Q - m^2_E)/M^2_1$,
$(m^2_D - m^2_L)/M^2_1$ and $(m^2_Q - m^2_U)/M^2_1$.
Here one could replace $M_1$ by any of the other two gaugino masses which simply would
amount in an overall rescaling. 
In Fig.~\ref{fig:ana} we show these 'invariants' in the leading-log
approximation at 1-loop order to demonstrate the principal behaviour
for seesaw type-II with a pair of ${\bf 15}$-plets and seesaw type-III
with three ${\bf 24}$-plets. From this one concludes that in principle
one has a handle to obtain information on the seesaw scale for given
assumptions on the underlying neutrino mass model, if universal boundary 
conditions are assumed.  For the type-I, i.e. singlets only, of
course $\Delta b_i=0$ and no change with respect to mSUGRA are
expected.  If, for example, the seesaw III model would be realized in
nature with three {\bf 24}-plets having similar masses around
$10^{13}$ GeV one could e.g.\ show that the corresponding ratios
cannot be obtained with one pair of {\bf 15}-plets in the seesaw II
model, thus excluding this possibility. However, taking the seesaw II
with two pairs of {\bf 15}-plets one would obtain similar ratios as in
this case the corresponding additional beta-functions at 1-loop would
be $\Delta b_i=14$, e.g.~nearly equal to our seesaw III model.

\begin{figure}[t]
  \centering
  \begin{tabular}{cc}
\includegraphics[width=0.48\textwidth]{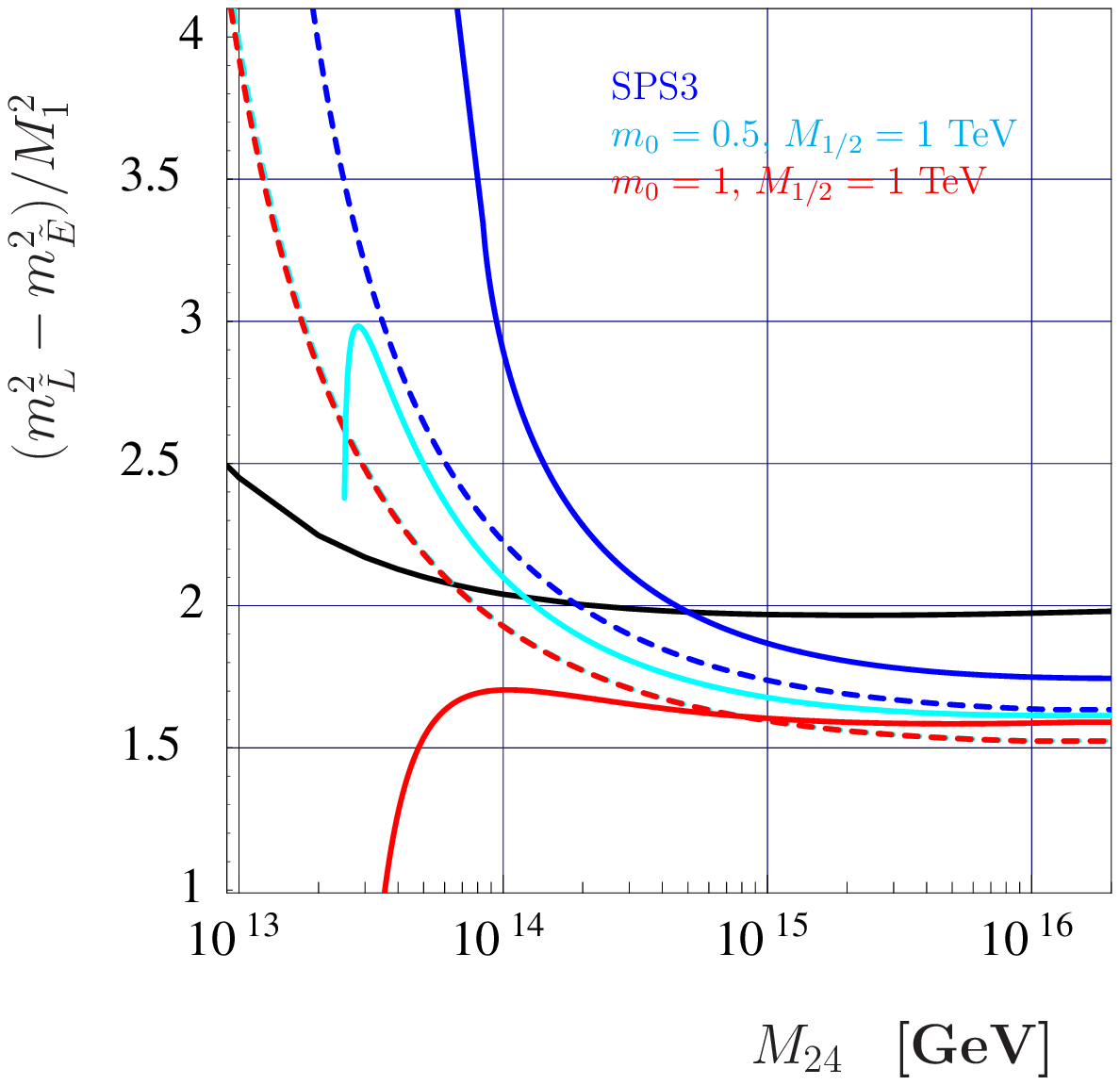}
\includegraphics[width=0.48\textwidth]{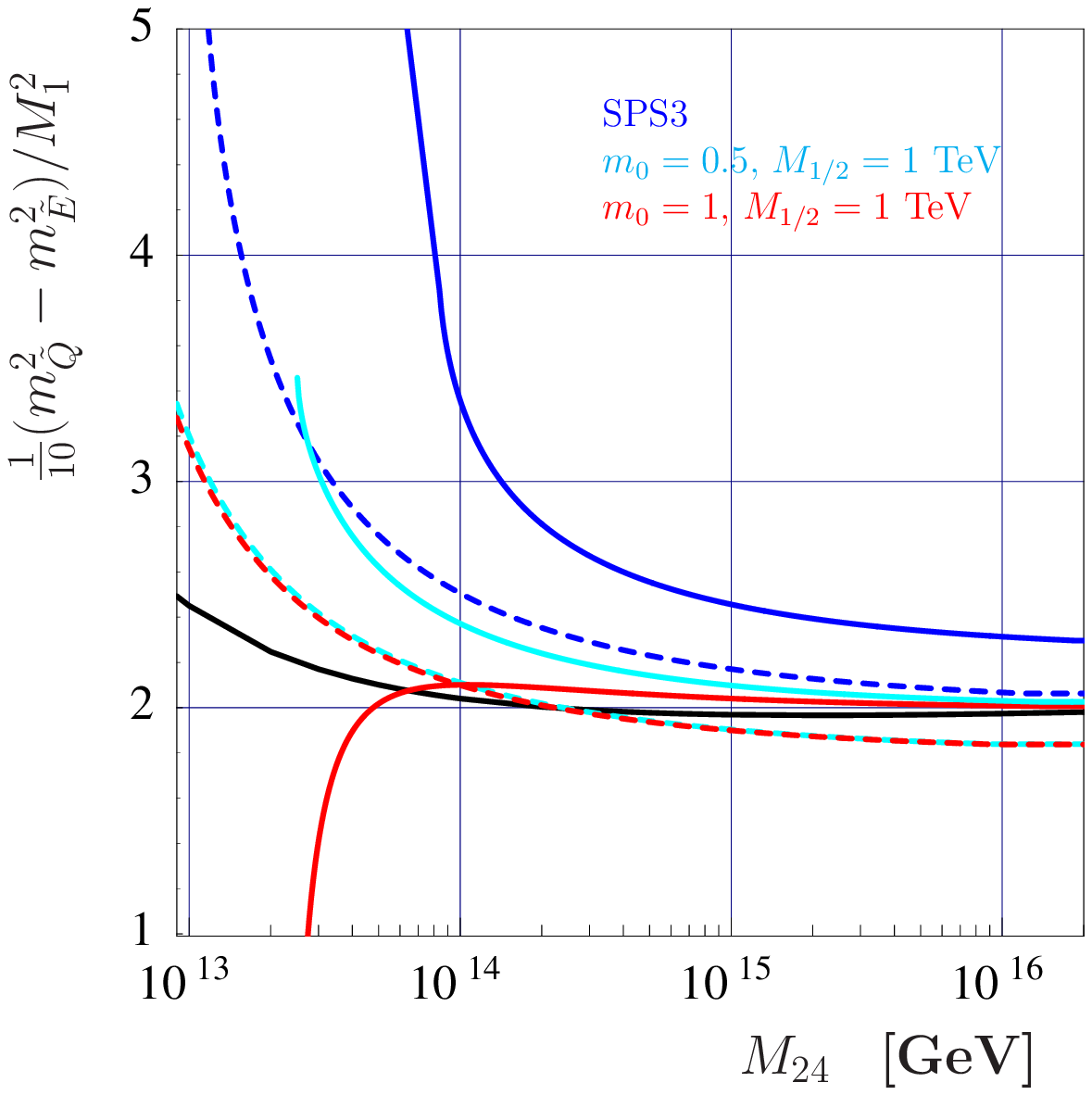}
\end{tabular}
\vspace{-5mm}
\caption{The limits of the invariants in seesaw type-III models. Left: $(m^2_L -
m^2_E)/M^2_1$, right $(m^2_Q - m^2_E)/M^2_1$. 
The blue lines are for SPS3, the light blue one for
$m_0=500$~GeV and $M_{1/2}=1$~TeV, and the red one for
$m_0=M_{1/2}=1$~TeV; full (dashed) lines are 2-loop (1-loop) results.
The black line is the analytical approximation, for comparison. }
  \label{fig:invfailed}
\end{figure}

The leading-log approximation gives only the general trend, but there 
is an important dependence on the SUSY point chosen.  In
Fig.~\ref{fig:invfailed} we show as illustration $(m^2_L -
m^2_E)/M^2_1$ and $(m^2_Q - m^2_E)/M^2_1$
for different mSUGRA points and at different loop orders: the dashed
lines are at 1-loop level whereas the solid ones are at 2-loop
level. The points considered are SPS3 \cite{Allanach:2002nj} with
$m_0=90$~GeV, $M_{1/2} =400$~GeV, $A_0=0$, $\tan\beta=10$, $\mu > 0$
and for the same values of $A_0$ and $\tan\beta$ two points with
$M_{1/2}=1$~TeV: $m_0=500$~GeV and $m_0=1$~TeV.
The black line shows for comparison the leading-log approximation.  We
observe that usually the approximation gets worse for lower values of
$M_{24}$ and this is even stronger at the 2-loop level which is a
consequence of the large coefficient in the beta functions at the
2-loop level, see e.g.~appendix \ref{app:BetaIII}. Nevertheless, one
sees that in general it gives the correct trend, but it might
even fail completely, e.g.~in the case of $M_{1/2}=m_0=1$~TeV.  The
reason for the drop around $M_{24} \simeq 3.5\times 10^{13}$ is that
the difference between the parameters goes to zero as can also be seen
from the right of Fig.~\ref{fig:mass1000}, see also discussion below. 

\begin{figure}[t]
  \centering
\includegraphics[width=0.48\textwidth]{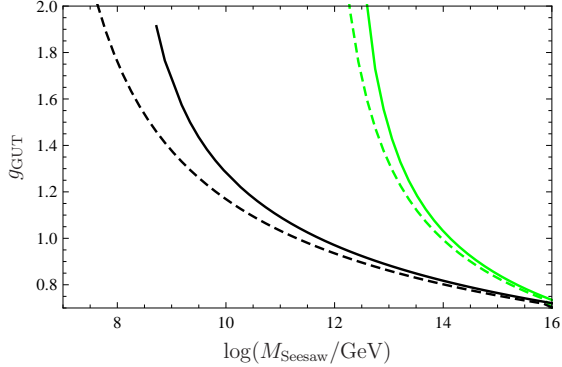}
\caption{Values of the gauge coupling at $M_{GUT}=2 \times 10^{16}$~GeV
as a function of the seesaw scale, black lines seesaw type-II and green
lines seesaw type-III with three {\bf 24}-plets with degenerate mass
spectrum; full (dashed) lines are 2-loop (1-loop) results. For the
calculation of the electroweak threshold the spectrum corresponds to
$m_0=M_{1/2}=1$~TeV, $A_0=0$, $\tan\beta=10$ and $\mu>0$.}
  \label{fig:landau}
\end{figure}

Last but not least we note that the use of the 2-loop RGEs leads to a
shift of $M_{GUT}$ from about 2$\times 10^{16}$ GeV for {\bf 24}-plet
mass of $10^{16}$ GeV to about 4$\times 10^{16}$ GeV for {\bf 24}-plet
mass of $10^{13}$ GeV, which is part of the differences between the
1-loop and 2-loop results in Fig.~\ref{fig:invfailed}.  Here $M_{GUT}$
is defined as the scale where the electro-weak couplings meet,
e.g.~$g_{U(1)} = g_{SU(2)}$.  This implies also that there is some
difference for the strong coupling which is, however, in the order of
5-10\% which can easily be accounted for by threshold effects of the
new GUT particles, e.g.~the missing members of the gauge fields and
the Higgs fields responsible for the breaking of the GUT group
\cite{Martens:2010nm}.  A second reason why the deviations between the
leading log calculation, the case of 1-loop and 2-loop RGEs gets
larger for smaller seesaw scale is that the increase of the beta
coefficients implies larger values of the gauge couplings at the GUT
scale.  This implies that one reaches a Landau pole for sufficiently
low values of the seesaw scale.  As an example we show in
Fig.~\ref{fig:landau} the value of the gauge coupling at $M_{GUT}=2
\times 10^{16}$~GeV as a function of the seesaw scale for type-II with
a pair of {\bf 15}-plets (black lines) and type-III with three
degenerate {\bf 24}-plets (green lines). In both cases the 2-loop RGEs
imply a larger gauge coupling for a fixed seesaw scale. One sees that
in case of type-II (type-III) in principle one could reach a seesaw
scale of about $10^{8}$~GeV ($10^{13}$~GeV).  However, we believe 
that we can no longer trust even the 2-loop calculation for such 
large values of the $g_i$, as the neglected higher order terms become 
more and more important. Especially, we should not trust the ``turn-over'' 
of the invariants in Fig.~\ref{fig:invfailed} for very low values of the 
seesaw scale, since the numerical calculation at these points is 
already very close to breaking down. 

We would also like to mention that, in the numerical 
calculation we find very often that one of the scalar masses squared, 
in particular staus and/or sbottoms, gets large negative values 
already for values of the seesaw scale larger than the Landau pole 
and thus we can not go to values of the seesaw scale as low as 
the examples shown in Fig.~\ref{fig:invfailed} in many SUSY points. 

\subsection{Lepton flavour violation in the slepton sector}
\label{sec:model_LFV}

>From a one-step integration of the RGEs one gets assuming mSUGRA boundary
conditions a first rough
estimate for the lepton flavour violating entries in the slepton
mass parameters:
\begin{eqnarray}
m^2_{L,ij} &\simeq& -\frac{a_k}{8 \pi^2 } 
 \left( 3 m^2_0 +  A^2_0 \right) 
 \left(Y^{k,\dagger}_N L Y^{k}_N\right)_{ij} \thickspace, \\
A_{l,ij} &\simeq& -a_k \frac{3}{ 16 \pi^2 }   A_0 
 \left(Y_e Y^{k,\dagger}_N L Y^{k}_N\right)_{ij} \thickspace,
\label{eq:LFVentriesI}
\end{eqnarray}
for $i\ne j$ in the basis where $Y_e$ is diagonal, 
$L_{ij} = \ln(M_{GUT}/M_i)\delta_{ij}$ and $Y^k_N$ is the additional Yukawa 
coupling of the type-$k$ seesaw at $M_{GUT}$ ($k=I, II, III$). We obtain
\begin{equation}
a_I=1\,\, , \,\, a_{II}=6 \,\, \mathrm{and} \,\,\, a_{III} = \frac{9}{5} \, .
\label{eq:LFVentriesII}
\end{equation}
Note, that in case of the type-II the matrix $L$ is degenerate and 
thus can be factored out. 
All models have in common that they predict negligible flavour
violation for the right-sleptons
\begin{eqnarray}
m^2_{E,ij} &\simeq& 0.
\end{eqnarray}
We know that these approximations work well only in case of the type-I models.
Nevertheless they give a rough idea on the relative size one has to expect
for the rare lepton decays $l_i \to l_j \gamma$ which very roughly scale
like
\begin{equation}
Br(l_i \to l_j \gamma) \propto \alpha^3 m_{l_i}^5
	\frac{| m^2_{L,ij}|^2}{\widetilde{m}^8}\tan^2\beta.
\label{eq:LLGapprox}	
\end{equation}
where $\widetilde{m}$ is the average of the SUSY masses involved in the loops.
Note, that for a given set of high scale parameters both, the different size
of the flavour mixing entries and the changed mass spectrum, play a role.

\section{Numerical results}
\label{sec:res}

In this section we present our numerical calculations. All results
presented below have been obtained with the lepton flavour violating
version of the program package SPheno \cite{Porod:2003um,SPheno}. The
RGEs of the seesaw II and seesaw III models have been calculated with
SARAH \cite{Staub:2008uz,Staub:2009bi,Staub:2010jh}. All seesaw
parameters are defined at  $M_{GUT}$ and as mentioned in the
previous section we require for models of type-II the boundary condition
  $Y_Z=Y_S=Y_T$ and $M_Z=M_S=M_T$  and in case of
type-III models $Y_N=Y_B=Y_W$ and $M_B=M_G=M_W=M_X$.
  We evolve the RGEs to the scale(s) corresponding
to the GUT scale values of the masses of the heavy particles. The RGE
evolution implies also a splitting of the heavy masses. We therefore
add at the corresponding scale the threshold effects due to the heavy
particles to account for the different masses. In case of type-III
models off-diagonal elements are induced in the mass matrices. This
implies that one has to go the corresponding mass eigenbasis before
calculating the threshold effects.  We use 2-loop RGEs everywhere
except stated otherwise.  In the appendix we give the necessary
ingredients on how to obtain them in the seesaw type-II and III models. The
analogous anomalous dimensions for the type-I model can be found in
\cite{Antusch:2002ek}.

Unless mentioned otherwise, we fit neutrino mass squared differences
to their best fit values \cite{Schwetz:2008er} and the angles to 
tri-bi-maximal (TBM)
values \cite{Harrison:2002er}. Our numerical procedure is as
follows. Inverting the seesaw equation, see eqs.~(\ref{eq:ssII}) and
(\ref{eq:mnu_seesawIII}), one can get a first guess of the Yukawa
couplings for any fixed values of the light neutrino masses (and
angles) as a function of the corresponding triplet mass for any fixed
value of the couplings.  This first guess will not give the correct
Yukawa couplings, since the neutrino masses and mixing angles are
measured at low energy, whereas for the calculation of $m_\nu$ we need
to insert the parameters at the high energy scale. However, we can use
this first guess to run numerically the RGEs to obtain the exact
neutrino masses and angles (at low energies) for these input
parameters. The difference between the results obtained numerically
and the input numbers can then be minimized in a simple iterative
procedure until convergence is achieved.  As long as neutrino Yukawas
are $\forall Y_{ij}< 1$ we reach convergence in a few steps.
However, in seesaw type-II and type-III the Yukawas run stronger than
in seesaw type-I, so our initial guess can deviate sizable from the
correct Yukawas, implying in general also more iterations until full 
convergence is reached. Since neutrino data requires at least one neutrino
mass to be larger than about $0.05$ eV, we do not find any solutions
for $M_T \gsim {\lambda_2} \times 10^{15}$ GeV and $M_{24} \gsim 8\times
10^{14}$ GeV, respectively.  In the latter case we have assumed that
all 24-plets have similar masses.  For sake of completeness we note
that one can also satisfy all neutrino data by giving one of the
24-plets a large mass in the order of $M_{GUT}$ or larger having a
model with effectively only two 24-plets.

\subsection{Lepton flavour violation}

We have seen in eq.~(\ref{eq:LLGapprox}) that rates for the lepton
flavour violating decays of $\mu$ and $\tau$ scale like the LFV
entries in the slepton mass squared matrix squared and inverse to the
overall SUSY mass to the power eight. From this one immediately
concludes the rates for the rare lepton decays are in general larger
in seesaw models of type-II and III than in type-I models for fixed
SUSY masses and seesaw scales except if one arranges for special
cancellations. 

\begin{figure}[t]
  \centering
  \begin{tabular}{cc}
  \includegraphics[width=0.45\linewidth]{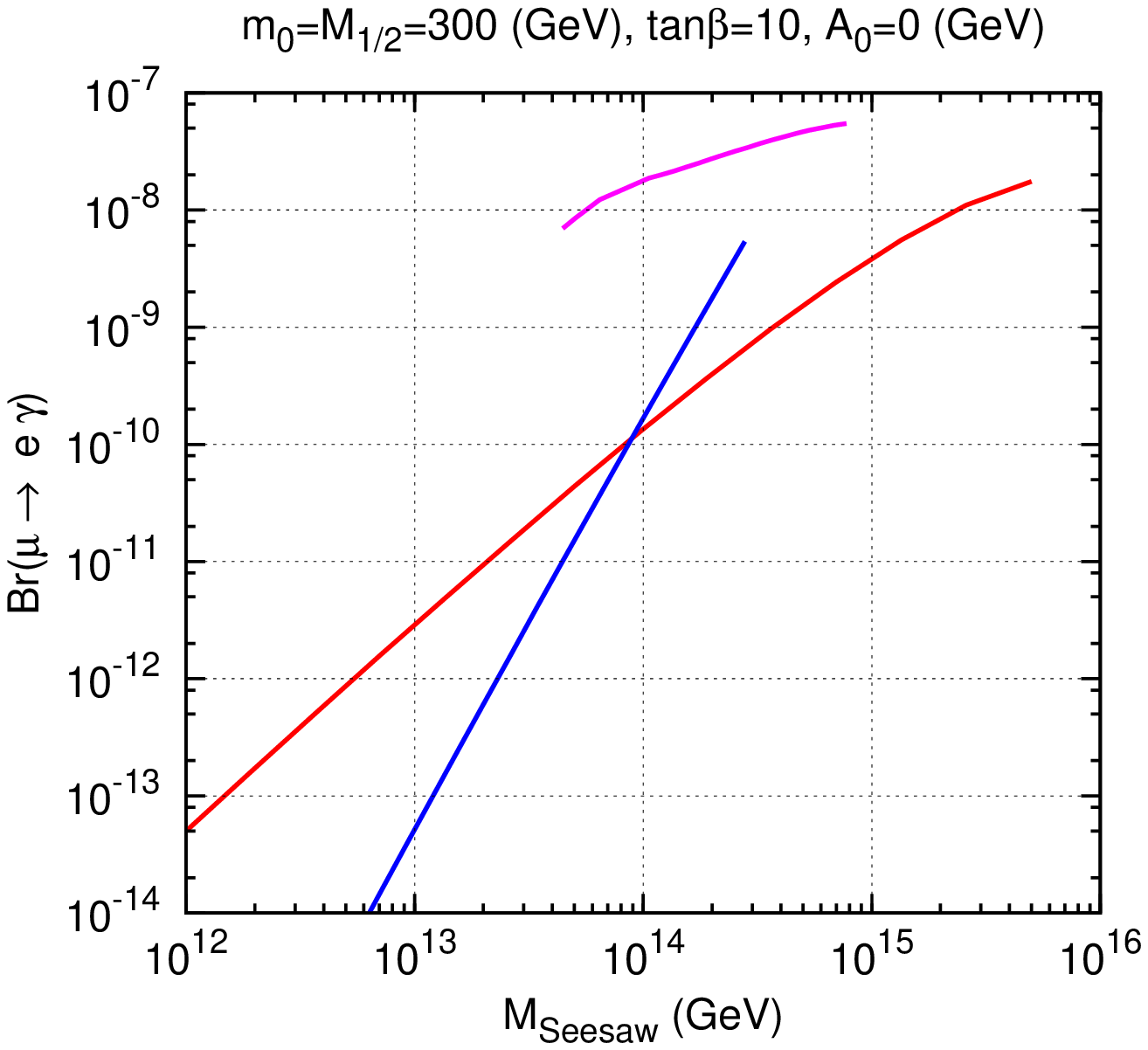}  &
  \includegraphics[width=0.45\linewidth]{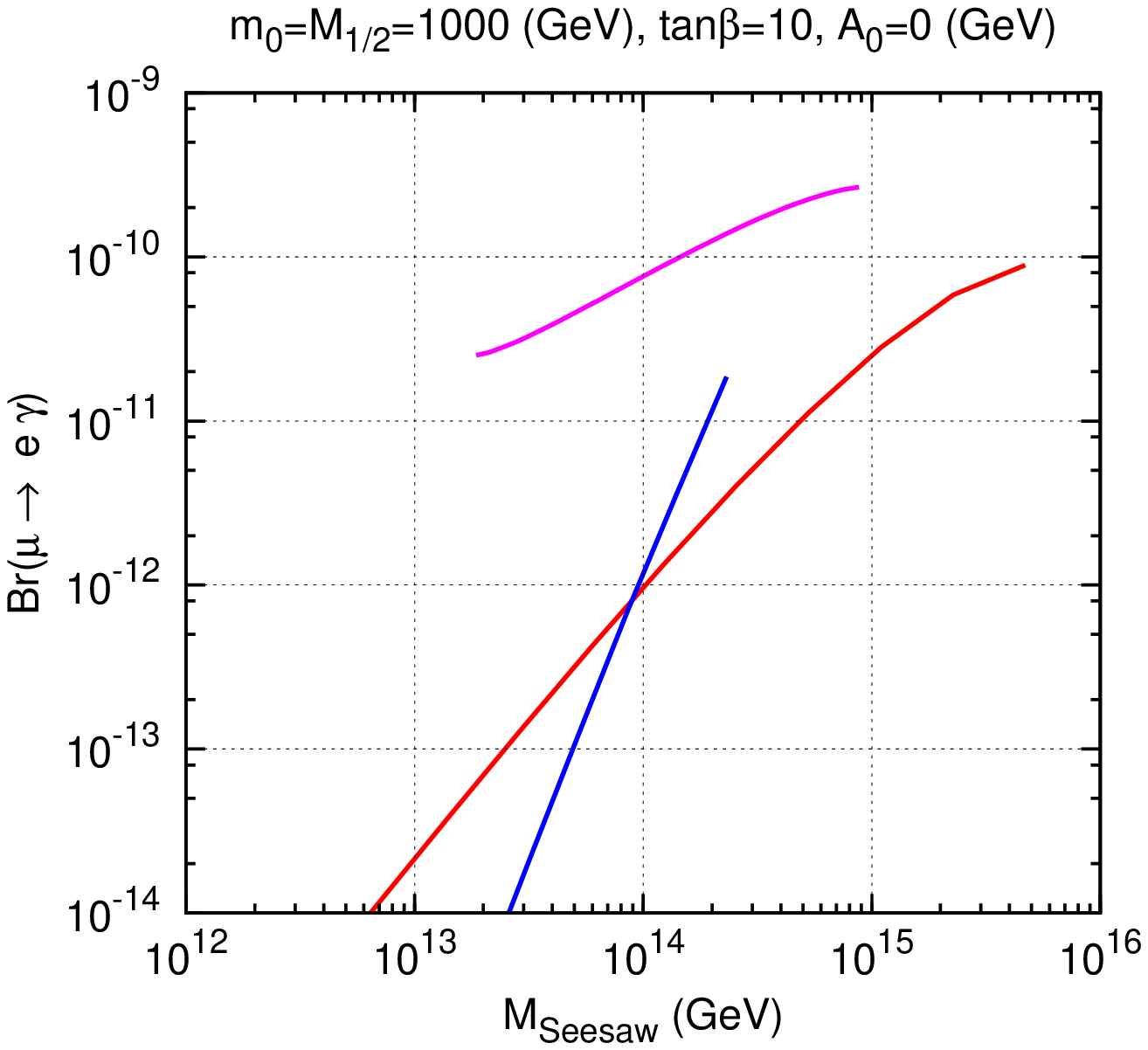}      
  \end{tabular}
  \caption{$Br(\mu \rightarrow e \gamma)$ as a function of the seesaw scale
   for seesaw type-I (red line), seesaw type-II (blue line)
   and seesaw type-III (magenta line). In case of type-I and type-III
   a degenerate spectrum has been assumed. On the left panel
  $m_0=m_{1/2}=300$ (GeV), on the right panel $m_0=m_{1/2}=1000$
  (GeV). In both cases we take $\tan\beta=10$, $A_0$=0 and $\mu>0$.}
  \label{fig:5a}
\end{figure}

Comparing the type-II with the type-III model one finds that LFV 
decays are larger for type-III, as shown for the case of $\mu\to e \gamma$
in Fig.~\ref{fig:5a}. From eqs.~(\ref{eq:LFVentriesI}) and 
(\ref{eq:LFVentriesII}), however, one would expect that type-II 
should have larger LFV. Numerically we find the opposite for two 
reasons. (i) $Br(l_i\to l_j \gamma)$ strongly depends on the SUSY 
masses, see eq. (\ref{eq:LLGapprox}) and type-III has a lighter spectrum 
than type-II (for the same mSUGRA input parameters). And (ii) 2-loop 
effects are very important in type-III, due to the large coefficients, 
in general leading to large flavor violating soft SUSY breaking parameters. 

In Fig.~\ref{fig:5a} we compare $Br(l_i\to l_j \gamma)$ for the three 
seesaw models taking degenerate seesaw spectra in case of type-I and 
type-III.   Note that in case of seesaw type-III 
we can only show a relatively short interval for the seesaw scale
which is mainly due to two reasons: (i) for scales below approximately
$10^{13}$ GeV the gauge couplings get large at $M_{GUT}$ as a
consequence of the large beta functions and, thus, perturbation theory
breaks down. (ii) One encounters negative mass squares for the
scalars, in particular for the lighter stau and/or lighter
sbottom. The latter point is also the reason why the possible range is
larger in case of the larger soft SUSY breaking parameters.

\begin{figure}[t]
  \centering
\includegraphics[width=0.45\linewidth]{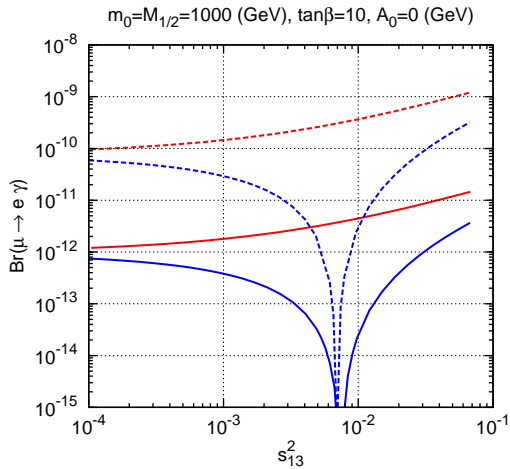}  
  \caption{$Br(\mu \rightarrow e \gamma)$ versus $s_{13}^2$ for
    $m_0=M_{1/2}=1000$ GeV, $\tan\beta=10$, $A_0=0$ GeV and $\mu>0$,
    for seesaw type-I (solid lines) and seesaw type-III (dashed
    lines), for $M_{\rm Seesaw} =10^{14}$ GeV. The curves shown are
    for 2 values of the Dirac phase: $\delta=0$ (red) and $\delta=\pi$
    (blue), both for normal hierarchy.}
  \label{fig:3a}
\end{figure}

\begin{table}[t]
\begin{center}
\begin{tabular}{c||c|c|c|c|c|c|c|c} 
$m_0$ & $m_{\tilde \chi^0_1}$ & $m_{\tilde \chi^+_1}$ & $m_{\tilde \chi^+_2}$ & $m_{\tilde g}$
 & $m_{\tilde \tau_1}$ & $m_{\tilde e_R}$  & $m_{\tilde e_L}$ &   $m_{\tilde t_1}$ \\ \hline
500 &  178 & 333 & 617 & 1029 & 535 & 543 & 600 & 772 \\ 
1000 &  180 & 338 & 642 & 1057 & 1008 & 1020 & 1043 & 925 \\ 
\end{tabular}
\end{center}
\caption{Examples masses in GeV for
    $M_{1/2}=1000$ GeV, $\tan\beta=10$, $A_0=0$ GeV and $\mu>0$,
    for seesaw  type-III  for a degenerate seesaw spectrum with $M_{24} =10^{14}$ GeV.}
    \label{tab:massIII}
\end{table}

\begin{figure}[t]
  \centering
  \begin{tabular}{cc}
\includegraphics[width=0.45\linewidth]{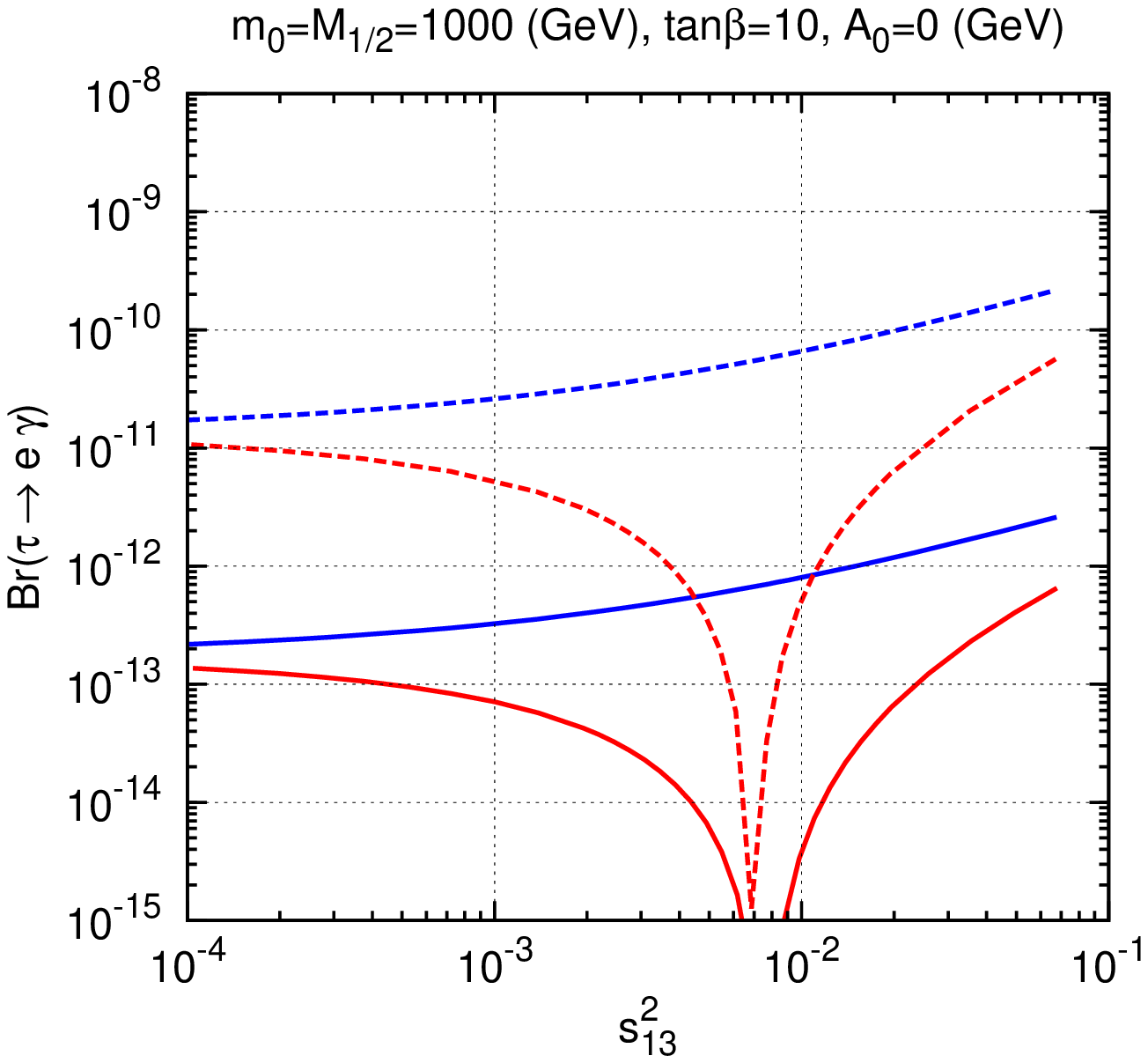}
&
\includegraphics[width=0.45\linewidth]{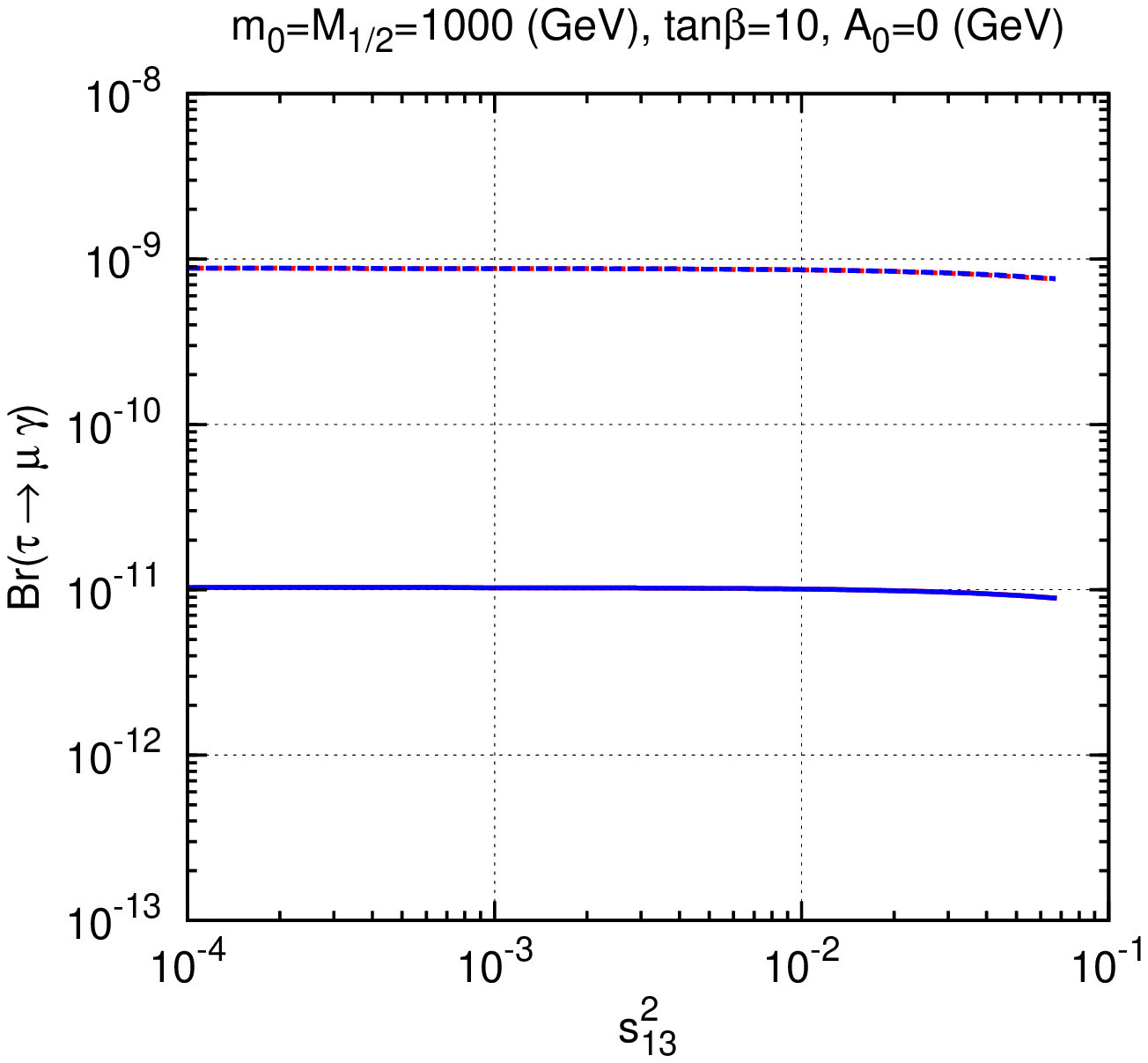}  
  \end{tabular}
  \caption{$Br(\tau \rightarrow e \gamma)$ versus $s_{13}^2$ (left) and 
    $Br(\tau \rightarrow \mu \gamma)$ versus $s_{13}^2$ (right)  for
    $m_0=M_{1/2}=1000$ GeV, $\tan\beta=10$, $A_0=0$ GeV and $\mu>0$,
    for seesaw type-I (solid lines) and seesaw type-III (dashed
    lines), for $M_{\rm Seesaw} =10^{14}$ GeV. The curves shown are
    for $\delta=0$ (red) and $\delta=\pi$
    (blue) for normal hierarchy.}
  \label{fig:3b}
\end{figure}

The values for $Br(\mu\to e\gamma)$ in Fig.~\ref{fig:5a} are larger 
than the current experimental bound \cite{Amsler:2008zzb}, so one might 
worry if in case of type-III models only SUSY spectra beyond the reach 
of the LHC are allowed. (Note, that even for the examples shown the masses
of the sfermions are already in the range of several hundred GeVs as
can be seen from table~\ref{tab:massIII}.) Indeed we find that by
putting generic Yukawa couplings which are able to explain neutrino
data one needs a heavy spectrum to be consistent with bounds on
the rare lepton decays. However, this is strictly true only for the 
TBM angles and $R=${\bf 1}. Accidental cancellations due to different contributions 
to the flavor violating soft masses and thus to the rare lepton decays are 
possible in type-III (and in type-I). As an example we show in 
Fig.~\ref{fig:3a} $Br(\mu\to e\gamma)$ as a function of the reactor 
angle $s^2_{13}$ for different values of the Dirac phase $\delta$. For comparison 
we also show the calculation for a type-I model. For $\delta = \pi$ 
there is a range of $s^2_{13}$ where this branching ratio is below the 
experimental constraint. 

At first glance this seem to require some fine-tuning of the underlying 
parameters. However, one can look at this from a different perspective: 
Assume that the MEG collaboration has found a non-vanishing value for 
$Br(\mu\to e\gamma)$ and from LHC data one has found that the spectrum 
is consistent with the type-III seesaw model. For a fixed $R$-matrix, e.g.~$R$={\bf 1}
 one would 
obtain in this case a relation between $s^2_{13}$ and $M_{24}$. This can be exploited 
to put a bound on $M_{24}$ or even to determine it depending on the 
outcome of measurements of reactor angle and, thus, the model assumptions can be
tested.
In Fig.~\ref{fig:3b} we show the corresponding rare tau
decays.  Note that also for $\tau\to e \gamma$ such a cancellation
exists  in principle 
but the corresponding range is excluded by $\mu\to e\gamma$. In contrast 
$\tau\to \mu \gamma$ is insensitive to the reactor angle
 and should be measurable in the near future.

\begin{figure}[t]
  \centering
  \includegraphics[width=0.45\linewidth]{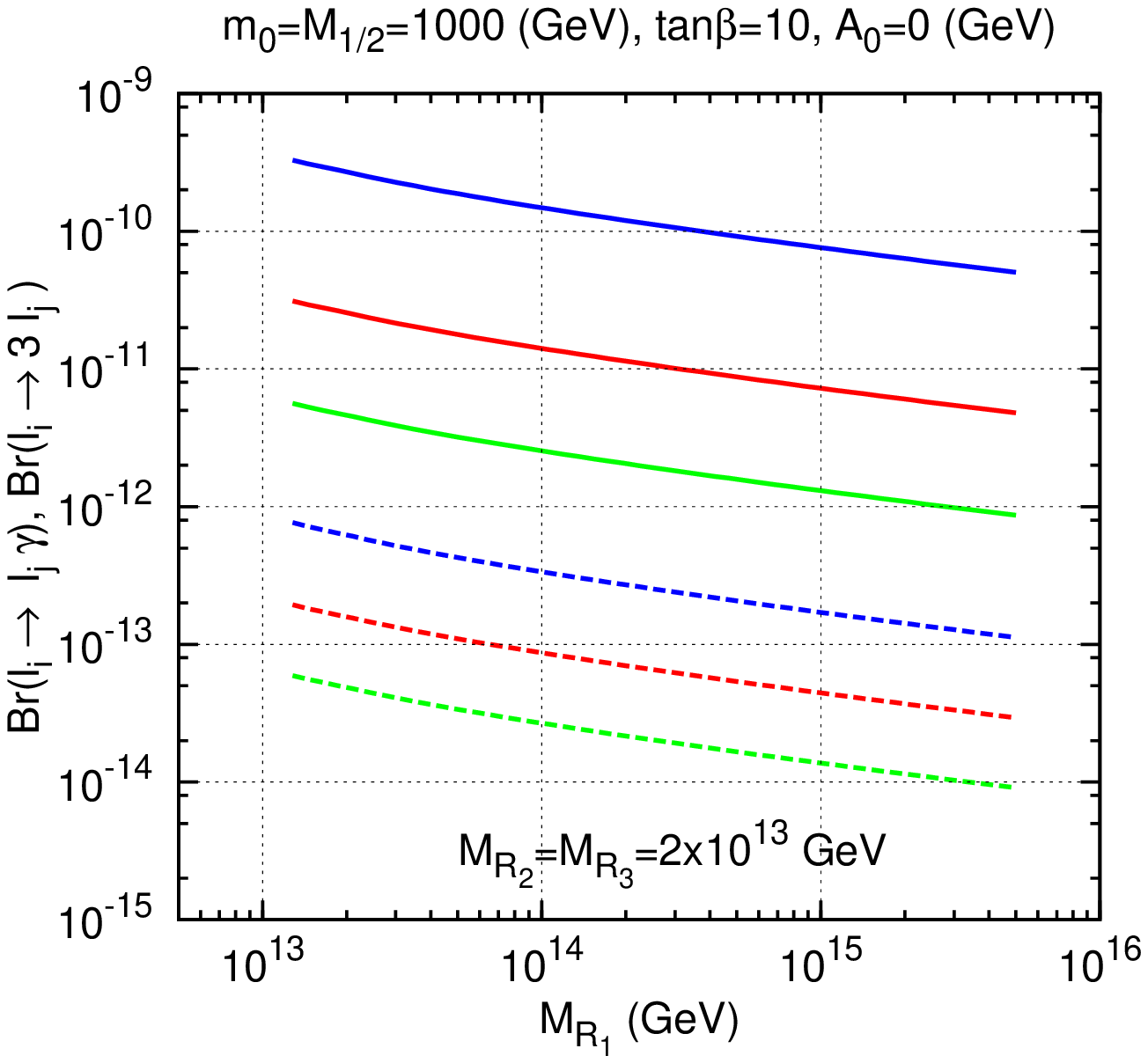}  
  \includegraphics[width=0.45\linewidth]{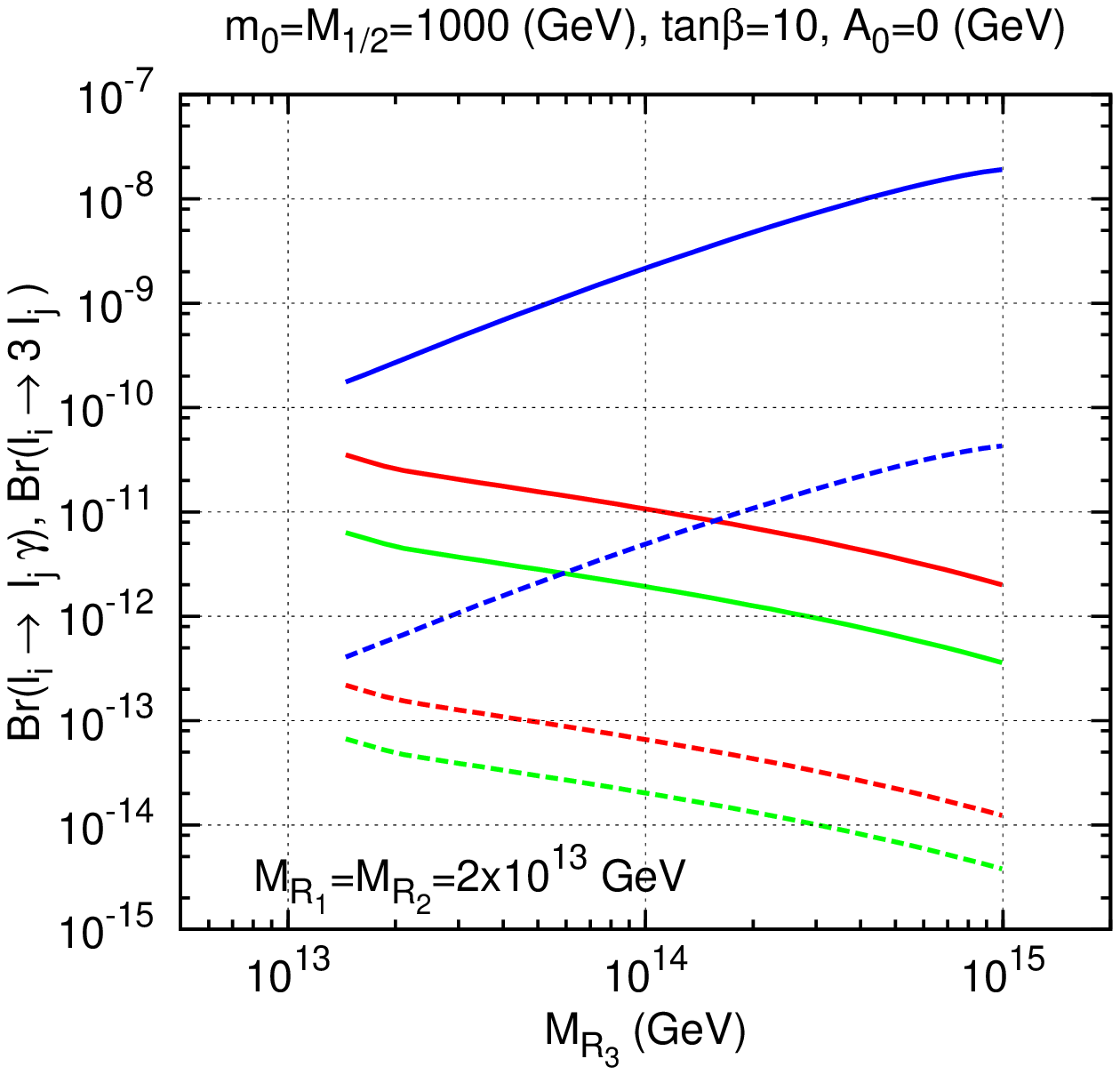}  
  \caption{Branching ratios for $l_i \to l_j \gamma$ (solid lines) and $l_i \to
    3l_j$ (dashed lines) versus the seesaw scale for $\tan\beta=10$, $\mu>0$, $A_O=0$
    GeV, $M_{1/2}=m_0=1000$ GeV. On the left panel we scan on $M_{R_1}$
    with $M_{R_2}=M_{R_3}=2\times10^{13}$ GeV while on the right panel we scan on 
    $M_{R_3}$
    with $M_{R_1}=M_{R_2}=2\times10^{13}$ GeV. The color code is red for $\mu\to e
    \gamma$ or $\mu \to 3 e$, blue for $\tau\to \mu \gamma$ or $\tau
    \to 3 \mu$ and green for $\tau\to e \gamma$ or $\tau \to 3 e$.}
  \label{fig:NonDeg1}
\end{figure}

Up to now we have assumed that the seesaw spectrum is nearly
degenerate which is of course a strong assumption. We show in
Fig.~\ref{fig:NonDeg1} two examples where we keep in each case two
masses fixed and vary the third one. Note, that in contrast to SUSY
particles the indices of the heavy particles are generation indices
and do not correspond to a particular mass ordering, e.g.~$M_{R_2}$ corresponds
to the 'solar neutrino scale' and $M_{R_3}$ to the 'atmospheric neutrino scale'. 
 In case that the
mass of the first generation state is varied, e.g.~the left plot of
this figure, one finds a decrease of the branching ratios with
increasing seesaw mass $M_{R_1}$.  This is mainly 
caused by an increase of
the SUSY spectrum while at the same time neutrino physics is only
affected mildly requiring only a light increase of the corresponding
Yukawa couplings to obtain the correct neutrino masses.  If, on the
other hand, the mass $M_{R_3}$ of the third generation seesaw particles is
increased on needs also a sizable increase of the Yukawa couplings to
obtain the correct neutrino mass difference squared for the
atmospheric sector. This leads to the observed behaviour that the
branching ratios for $\tau \to \mu \gamma$ and $\tau \to 3 \mu$
increases while the other ones decrease.

\subsection{Dark Matter}

\begin{figure}[t]
  \centering
  \begin{tabular}{cc}
  \includegraphics[width=0.45\linewidth]{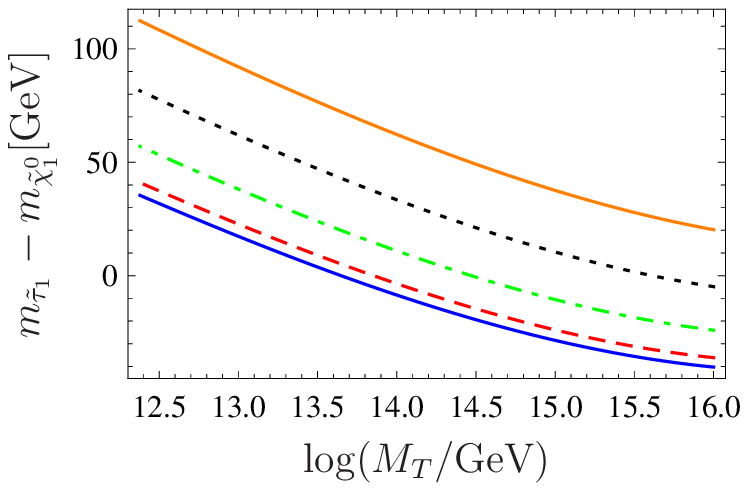}&
  \includegraphics[width=0.45\linewidth]{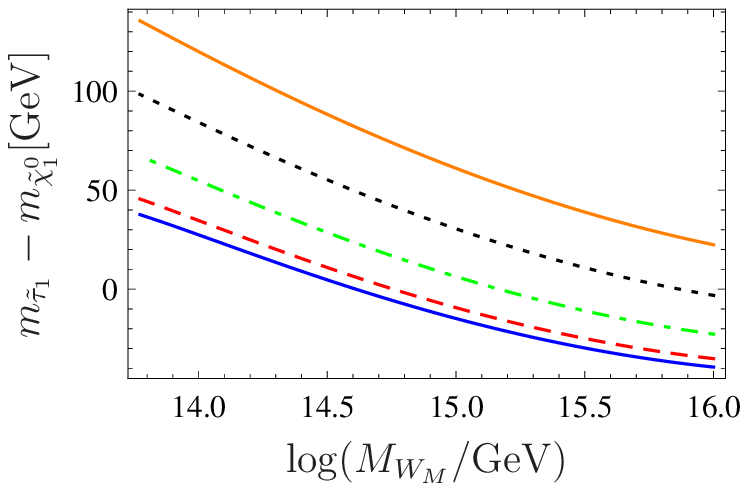} \\
  \includegraphics[width=0.45\linewidth]{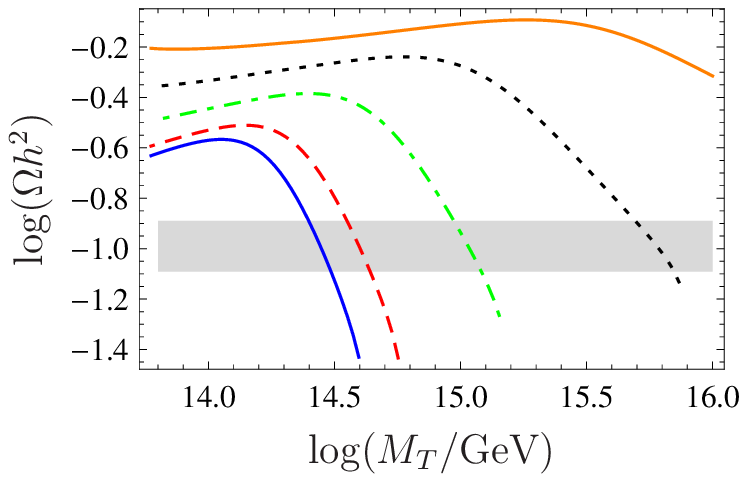}&
  \includegraphics[width=0.45\linewidth]{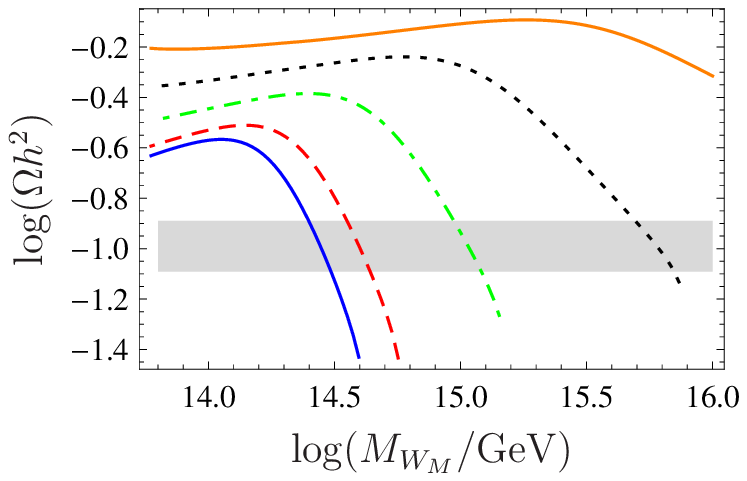} 
\end{tabular}
   \caption{Difference between the masses and the lightest stau and the lightest
    neutralino (upper row) as well as the corresponding $\Omega h^2$ (lower row)
    as a function of the seesaw scale. The left (right) plots are for seesaw type-II
    (III). A degenerate seesaw spectrum has been assumed in case of seesaw type-III.
    $M_{1/2}= 800$ GeV, $A_0=0$, $\tan\beta=10$ and $\mu>0$. The lines correspond to
    full blue line $m_0=0$, red dashed line $m_0=50$~GeV, green dashed dotted line
     $m_0=100$~GeV, black dashed line $m_0=150$~GeV and orange full line $m_0=200$~GeV. The gray band shows the preferred range according to eq.~(\ref{eq:Relic3S}).}
   \label{fig:coannihilation}
\end{figure}

The changes in the spectrum induced by the new heavy states also impact 
on the predictions with respect to the relic density which we have 
calculated using the program \texttt{micrOMEGAs} \cite{Belanger:2006is}. 
As is well-known, within mSUGRA there are 4 regions in parameter 
space, in which the constraint from dark matter can be satisfied. 
These are (i) the bulk region; (ii) the stau co-annihilation region; 
(iii) the focus point line and (iv) the Higgs funnel. Below we will 
show usually the range of $\Omega h^2$ allowed at 3 $\sigma$ 
according to \cite{Amsler:2008zzb}
\begin{equation}
\label{eq:Relic3S}
0.081 \le \Omega h^2 \le 0.129 \,.
\end{equation}

In particular, the co-annihilation region is very sensitive to the
difference between the masses of the lightest stau and the lightest
neutralino. In Fig.~\ref{fig:coannihilation} we observe that this
difference depends strongly on the seesaw scale in both models. For a
fixed $M_{1/2}$ and $m_0$ lowering the seesaw scale increases this
mass difference, which then leads to a larger calculated $\Omega
h^2$. To compensate for this effect one needs to lower $m_0$, with the
value depending on the seesaw scale chosen. For certain seesaw scales
then $m_0$ needs to be lowered below $m_0=0$ and the co-annihilation
region disappears. In this region of parameter space both models
behave in a qualitatively similar way.  However, recall that spectra
run faster towards smaller masses in seesaw type-III.

\begin{figure}[t]
  \centering
  \begin{tabular}{ccc}
  \includegraphics[width=0.32\linewidth]{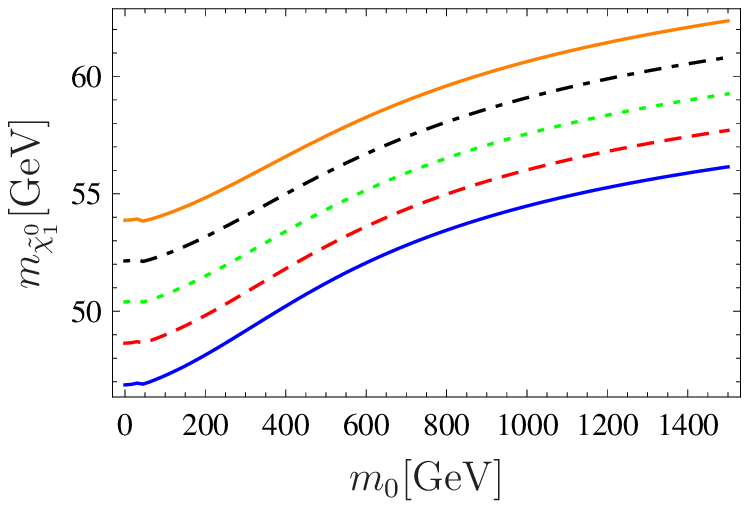}&
  \includegraphics[width=0.32\linewidth]{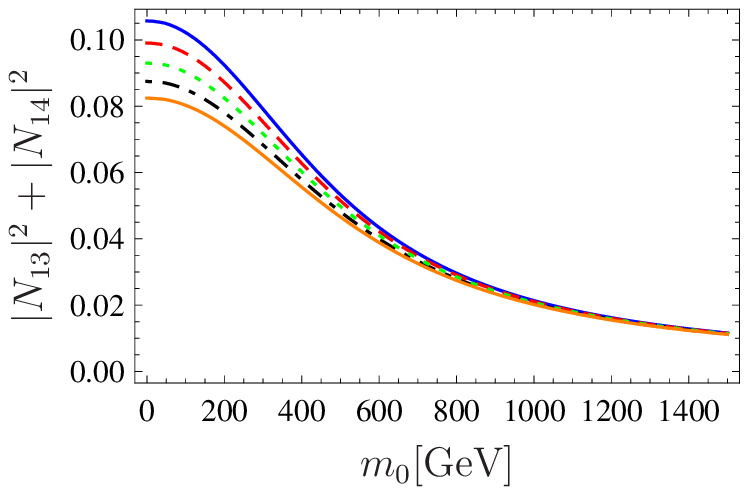} &
  \includegraphics[width=0.32\linewidth]{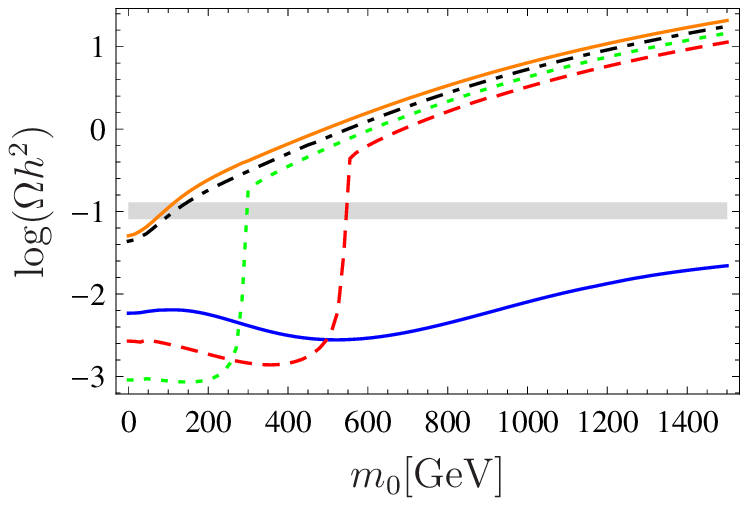} 
\end{tabular}
   \caption{Mass of the lightest neutralino (left plot), its higgsino content 
    (middle plot) and the corresponding  $\Omega h^2$ (right plot)
    as a function of $m_0$ for a seesaw type-II model with $M_T=10^{14}$ GeV,
     $m_{top}=171.2$ GeV,
    $A_0=0$, $\tan\beta=10$ and $\mu>0$. The lines correspond to
    full blue line $M_{1/2}=195$~GeV, red dashed line $M_{1/2}=200$~GeV, 
    green dashed dotted line
     $M_{1/2}=205$~GeV, black dashed line $M_{1/2}=210$~GeV and 
     orange full line $M_{1/2}=215$~GeV. The gray band  shows the range eq.~(\ref{eq:Relic3S}).}
\label{fig:focusII}
\end{figure}

\begin{figure}[t]
  \centering
  \begin{tabular}{ccc}
  \includegraphics[width=0.32\linewidth]{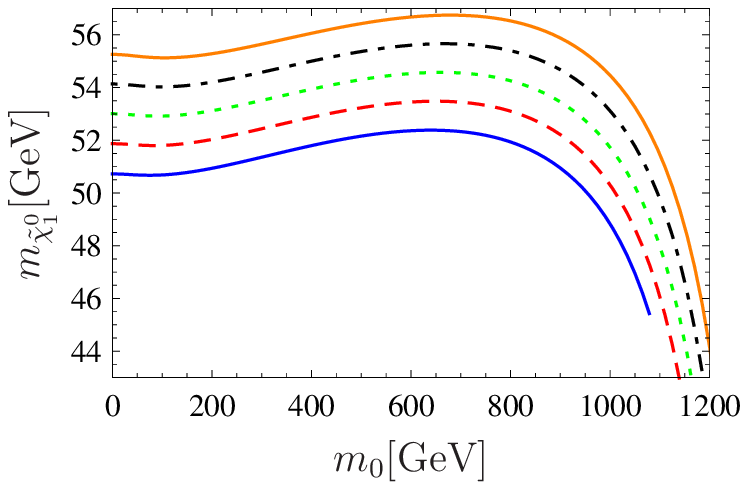}&
  \includegraphics[width=0.32\linewidth]{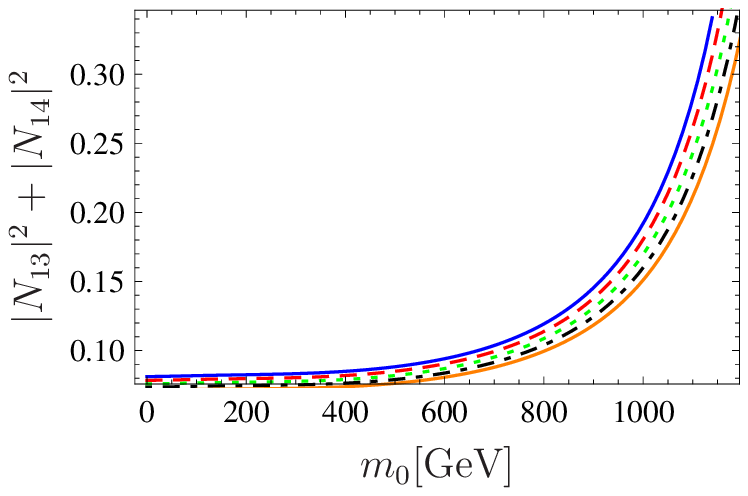} &
  \includegraphics[width=0.32\linewidth]{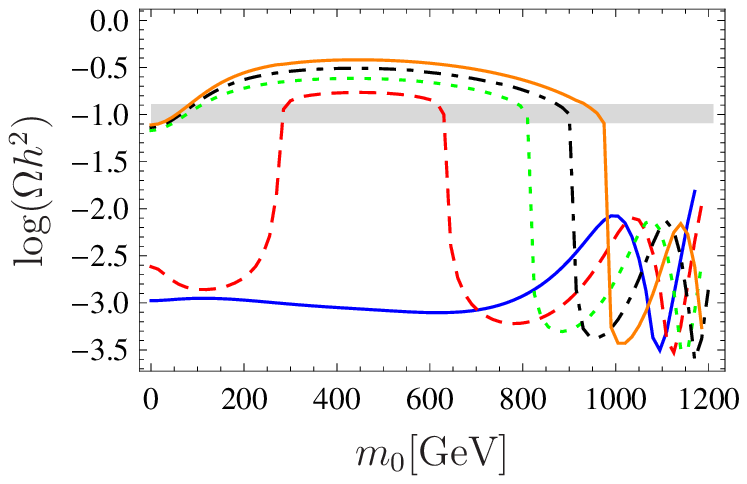} 
\end{tabular}
   \caption{Mass of the lightest neutralino (left plot), its higgsino content 
    (middle plot) and the corresponding  $\Omega h^2$ (right plot)
    as a function of $m_0$ for a seesaw type-III model with a degenerate
    seesaw scale $M_W=10^{14}$ GeV, $m_{top}=171.2$ GeV,
    $A_0=0$, $\tan\beta=10$ and $\mu>0$. The lines correspond to
    full blue line $M_{1/2}=400$~GeV, red dashed line $M_{1/2}=405$~GeV, 
    green dashed dotted line
     $M_{1/2}=410$~GeV, black dashed line $M_{1/2}=415$~GeV and 
     orange full line $M_{1/2}=420$~GeV. The gray band  shows the range eq.~(\ref{eq:Relic3S}).}
\label{fig:focusIII}
\end{figure}

Also the focus point region is very sensitive to the precise values of 
the input parameters. The focus point region appears in mSUGRA for 
large values of $m_0$ and small/moderate values of $M_{1/2}$ of 
the order of ${\cal O}(100)$ GeV, the exact value depending on $m_0$.
This can be seen in figs.~\ref{fig:focusII} and \ref{fig:focusIII} where 
we show $m_{\tilde \chi^0_1}$, the higgsino content $|N_{13}|^2+|N_{14}|^2$
and the corresponding $\Omega h^2$ as a function of $m_0$ for a fixed
seesaw scale $M_{T,W}=10^{14}$ GeV, $A_0=0$, $\tan\beta=10$, $\mu>0$
and various values of $M_{1/2}$. Note, that we take different values of 
$M_{1/2}$ for the two models in such a way that we obtain similar values 
for $m_{\tilde \chi^0_1}$. We find that both models behave differently 
in this region of parameter space, e.g.\ the higgsino content 
$|N_{13}|^2+|N_{14}|^2$ decreases (increases) with increasing values 
$m_0$ for seesaw type-II (type-III). However, also for type-II the higgsino 
content increases for increasing $m_0$ once we reach the multi-TeV range 
but we did not get correct electroweak symmetry breaking in case of 
multi-TeV values for $m_0$ in case of type-III models.
The increased higgsino content of the lightest neutralino leads to on increase (decrease)
of its couplings to the $Z$-boson and the light Higgs boson (to sfermions) resulting
in the observed dependence of $\Omega h^2$ for $m_0$ close to the 1-TeV region.

\begin{figure}[t]
  \centering
  \begin{tabular}{cc}
  \includegraphics[width=0.45\linewidth]{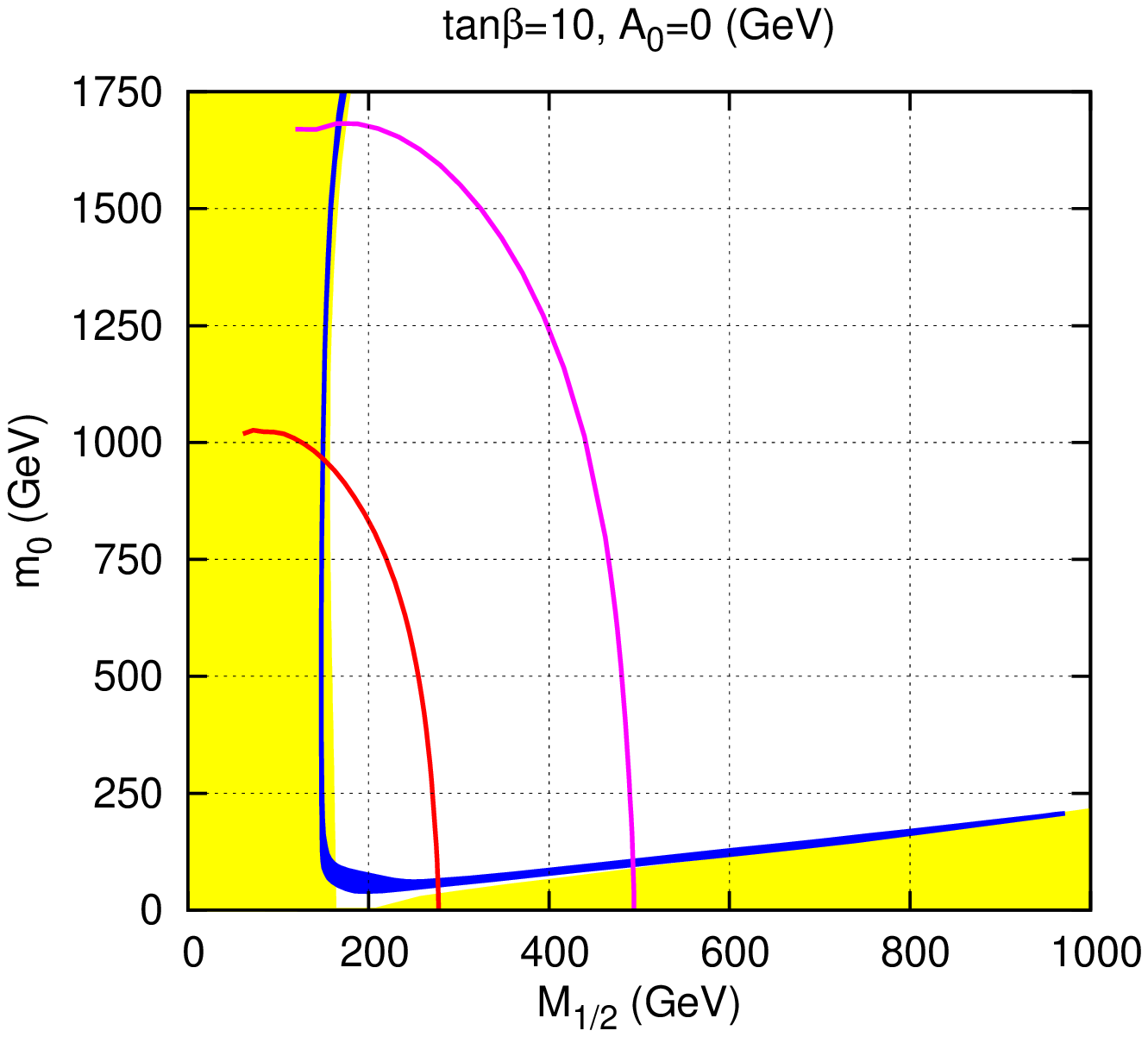}&
  \includegraphics[width=0.45\linewidth]{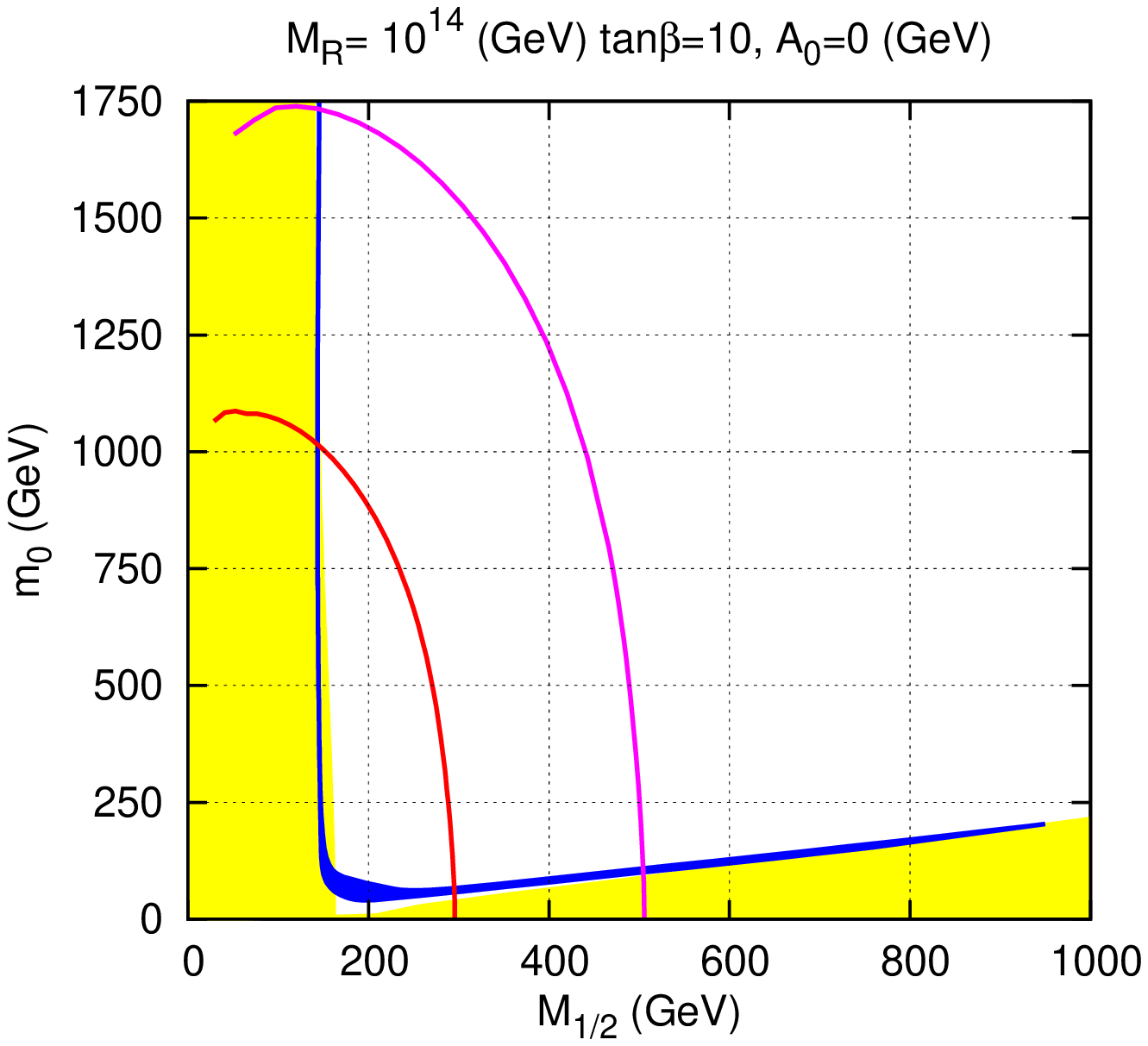}
\end{tabular}
   \caption{Dark matter allowed region (in blue) for mSUGRA (left
     panel) and for type-I seesaw (right panel). The parameters are
     $\tan\beta=10$, $A_0=0$, $\mu>0$ and
     $M_T=10^{14}$ GeV for  $m_{top}=171.2$ GeV. Also shown (in
     yellow) are the regions excluded by LEP (small values of
     $M_{1/2}$), and by LSP constraint (small values of $m_0$). Also
     shown are the Higgs boson mass curves for $M_h=110$ GeV (in red) and
     for $M_h=114.4$ GeV (in magenta).}
   \label{fig:2NewA}
\end{figure}

With these observations it is clear that the DM allowed regions will be 
shifted in the $m_0$-$M_{1/2}$ plane compared to the usual mSUGRA 
expectations. We fix in the following $m_{top}=171.2$~GeV, $\tan\beta=10$, 
$A_0=0$ and $\mu>0$ as well as the seesaw scale to $10^{14}$ GeV. For 
comparison we show in Fig.~\ref{fig:2NewA} the usual mSUGRA case without 
any heavy intermediate particles (left plot) as well as the case of a seesaw
type-I scenario (right plot).  The blue bands show the 3$\sigma$ range
according to \cite{Amsler:2008zzb} and we see the three usual regions:
the stau co-annihilation with a lighter stau mass close to the LSP mass
for $M_{1/2} \lsim 300$ GeV, the bulk region for moderate values of
$M_{1/2}$ and $m_0$ resulting in small sfermion masses as well as the
focus point region for $M_{1/2} \simeq$ 170 GeV and large values of
$m_0$. In addition, we show the lines corresponding to $M_h=110$~GeV
and 114 GeV. Note, that the theoretical uncertainty on $M_h$ is still 
of the order of 3-5 GeV \cite{Allanach:2004rh,Heinemeyer:2004xw}. Moreover, 
the value of the Higgs boson mass also depends strongly on $A_0$ and in particular 
for negative values of $A_0$ one can easily increase the value of $M_h$ 
while the DM allowed regions hardly change. 

The  part of parameter space most affected is the one at large $m_0$. 
Since in mSUGRA $\mu$ is calculated from the requirement of correct
electroweak symmetry breaking, $\mu$ changes rapidly in this region. 
With the Higgsino content in the lightest neutralino changing rapidly 
as a function of $\mu$, this region is then very sensitive to any 
changes of parameters. Since the $Y_\nu$ also impacts on the
running of the Higgs mass parameters and thus slightly affects the
value predicted for $\mu$, some small changes are found relative to mSUGRA 
here.  Note, however, that this region is highly 
constrained by the lower bound on the lightest chargino mass 
of the order of 103 GeV~\cite{LEPSUSYWG}.

\begin{figure}[t]
  \centering
  \begin{tabular}{cc}
  \includegraphics[width=0.45\linewidth]{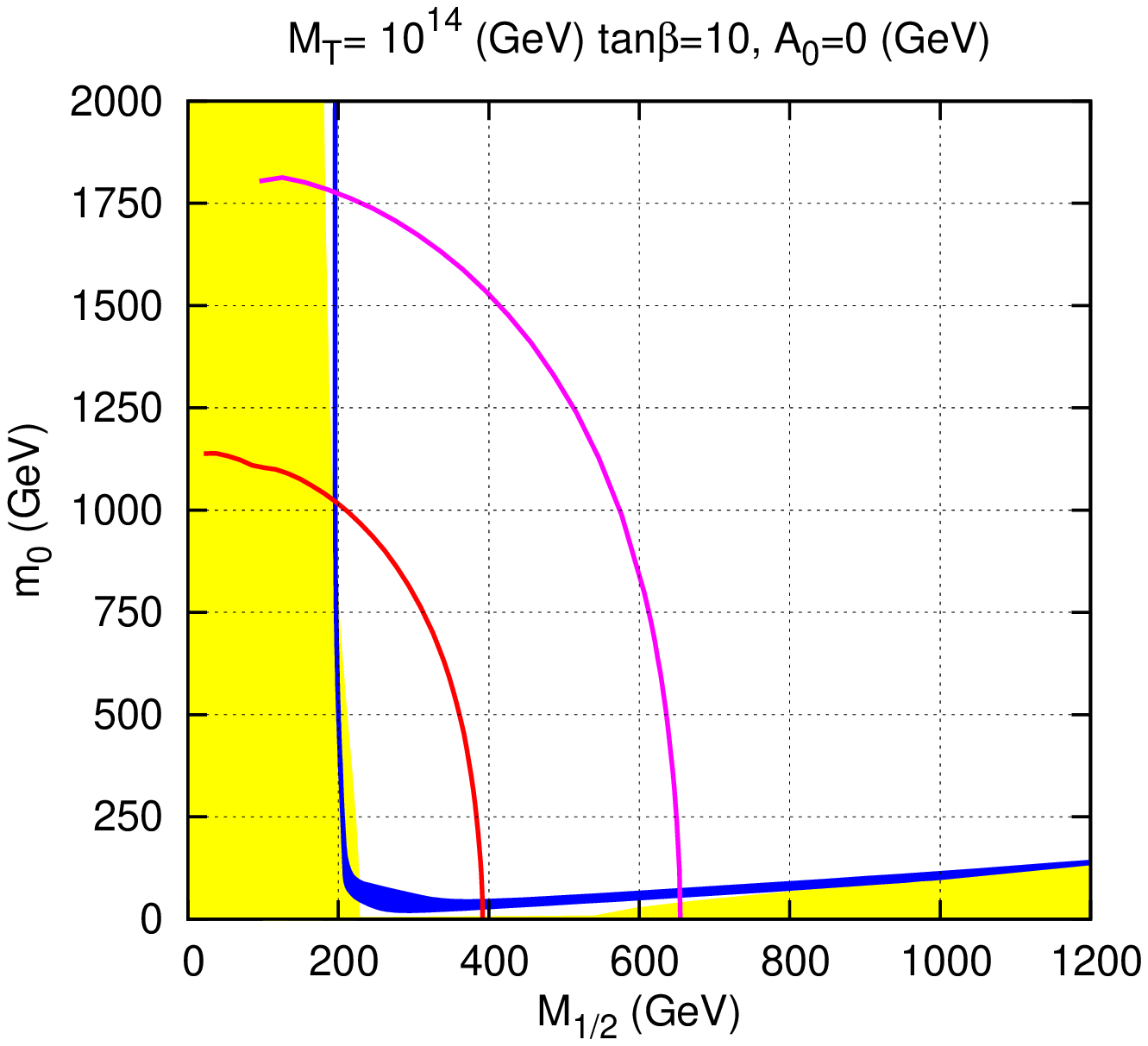}&
  \includegraphics[width=0.45\linewidth]{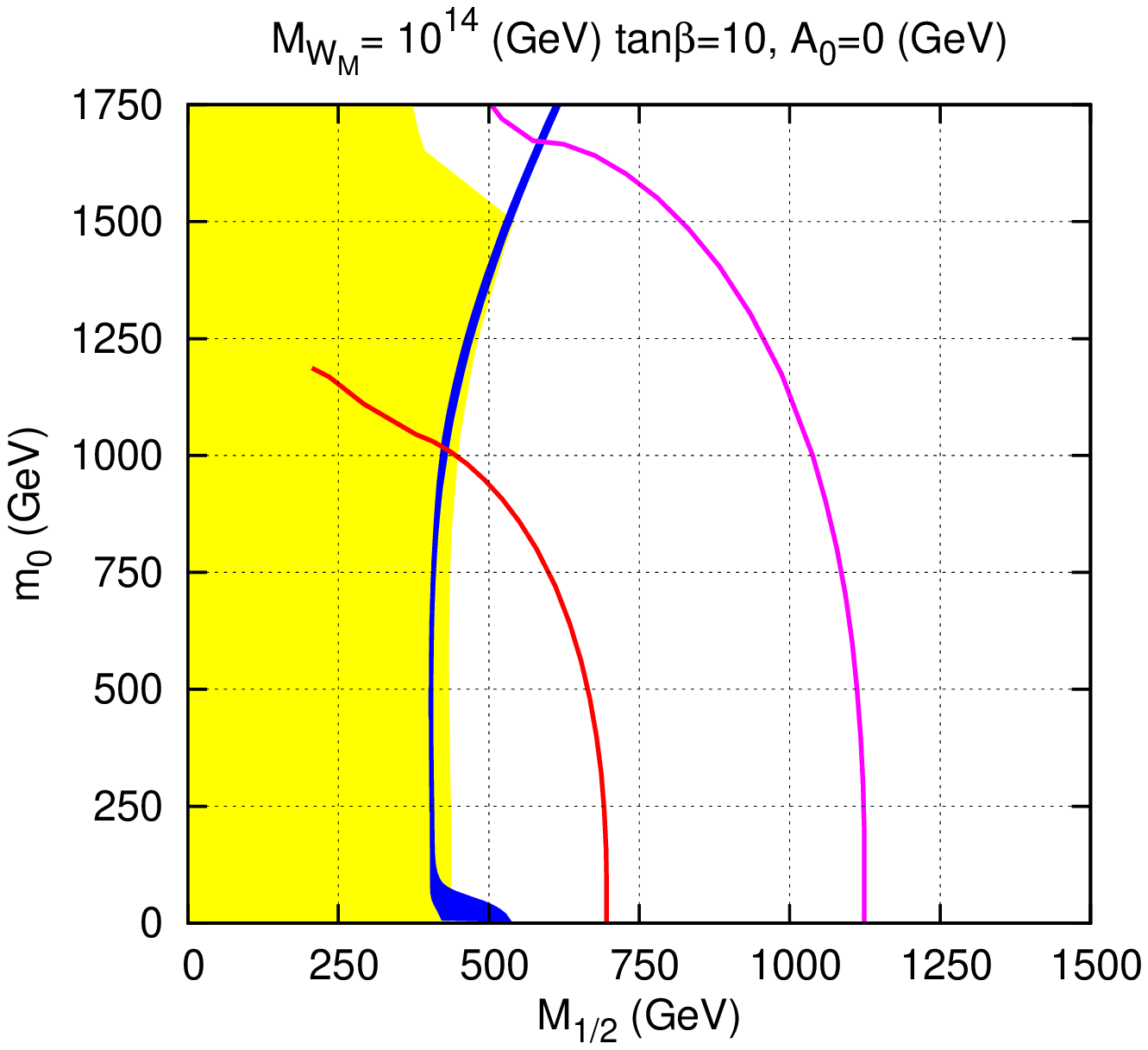}
\end{tabular}
   \caption{Like in Fig.~\ref{fig:2NewA} but for seesaw type-II (left
     panel) and type-III (right panel).}
   \label{fig:2NewB}
\end{figure}

\begin{figure}[t]
  \centering
  \begin{tabular}{cc}
  \includegraphics[width=0.45\linewidth]{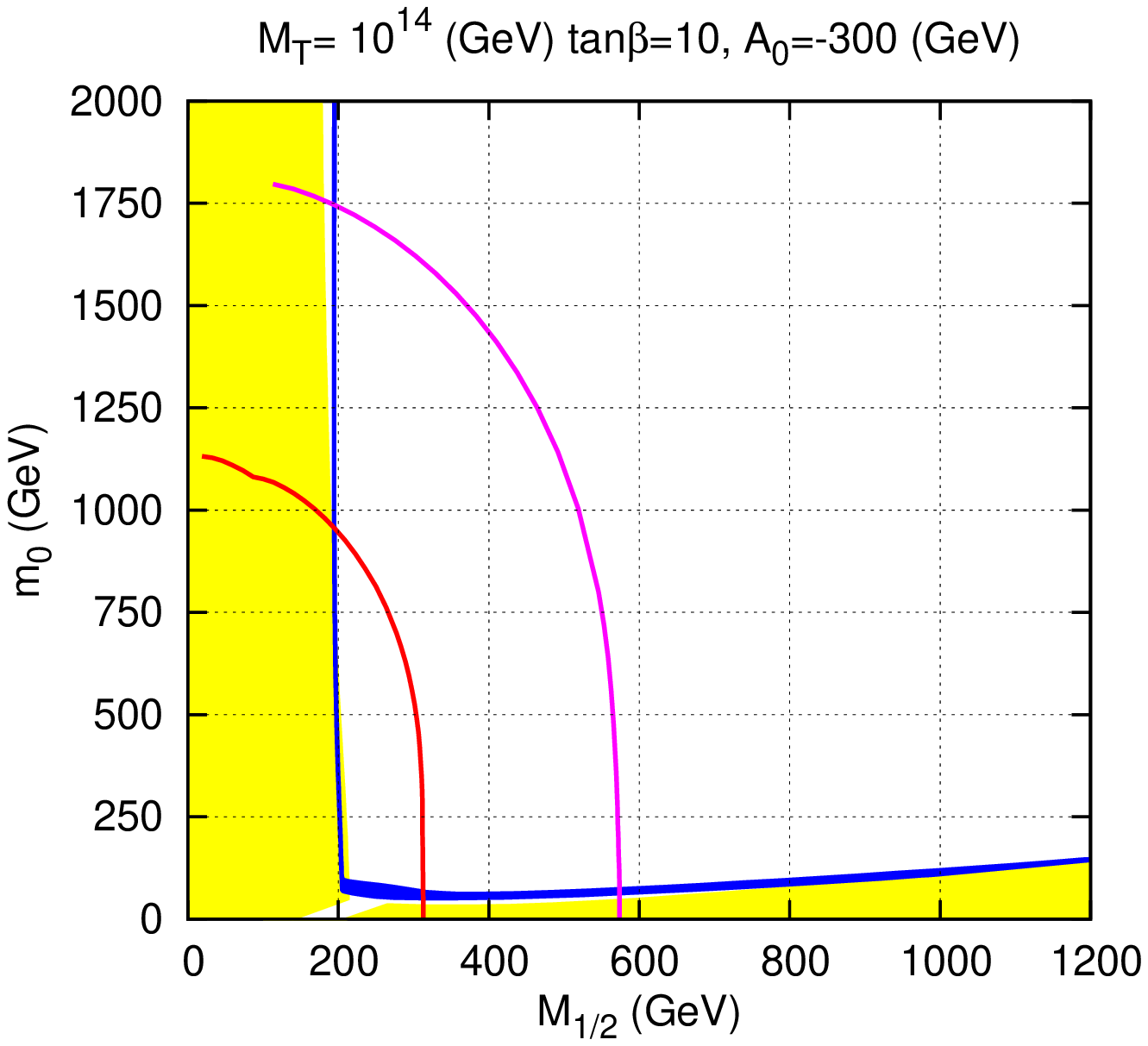}&
  \includegraphics[width=0.45\linewidth]{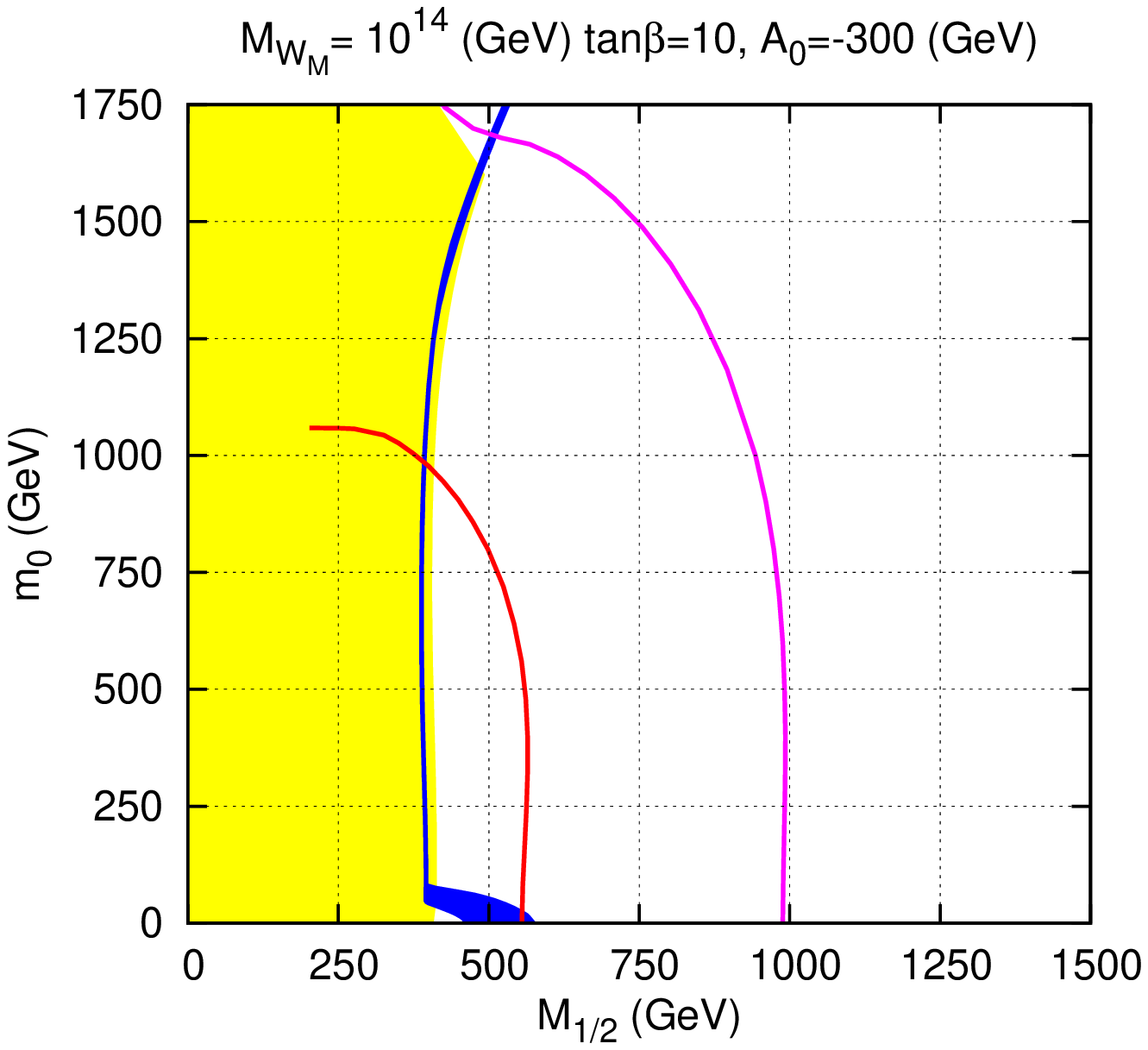}
\end{tabular}
   \caption{Like in Fig.~\ref{fig:2NewB} but for $A_0=-300$. Seesaw
     type-II (left panel) and type-III (right panel).}
   \label{fig:2NewC}
\end{figure}

In case of the other two seesaw models the shift of the allowed
regions is much more pronounced, as discussed above. In Figs.~\ref{fig:2NewB}
 and \ref{fig:2NewC} we show to regions for type-II
(left plot) and type-III (right plot) and two different values for
$A_0$. As claimed above, the Higgs mass bounds gets shifted
significantly while the DM allowed regions are hardly affected. As expected
the effects are much more pronounced in case of type-III as the
effects of the heavy particles on the spectrum is much stronger. Note,
that in particular the bending of the allowed region for large $m_0$
is due to the changed higgsino content as discussed in case of
figs.~\ref{fig:focusII} and \ref{fig:focusIII}. Moreover, the case of
stau co-annihilation is not viable anymore in case of the type-III
model already for this value of the seesaw scale. For completeness 
we mention that for the type-II the stau co-annihilation region disappears 
(below $M_{1/2}=1500$ GeV) for $M_T \lsim 10^{13}$ GeV. For completeness
we note that the results here differ slightly from the ones of our
previous work \cite{Esteves:2009qr} because (i) of the corrections
of the 1-loop
RGEs of ref.~\cite{Rossi:2002zb} by \cite{Borzumati:2009hu}
 and (ii) the complete set of 2-loop RGEs are now used.

\begin{figure}[t]
  \centering
  \begin{tabular}{cc}
  \includegraphics[width=0.45\linewidth]{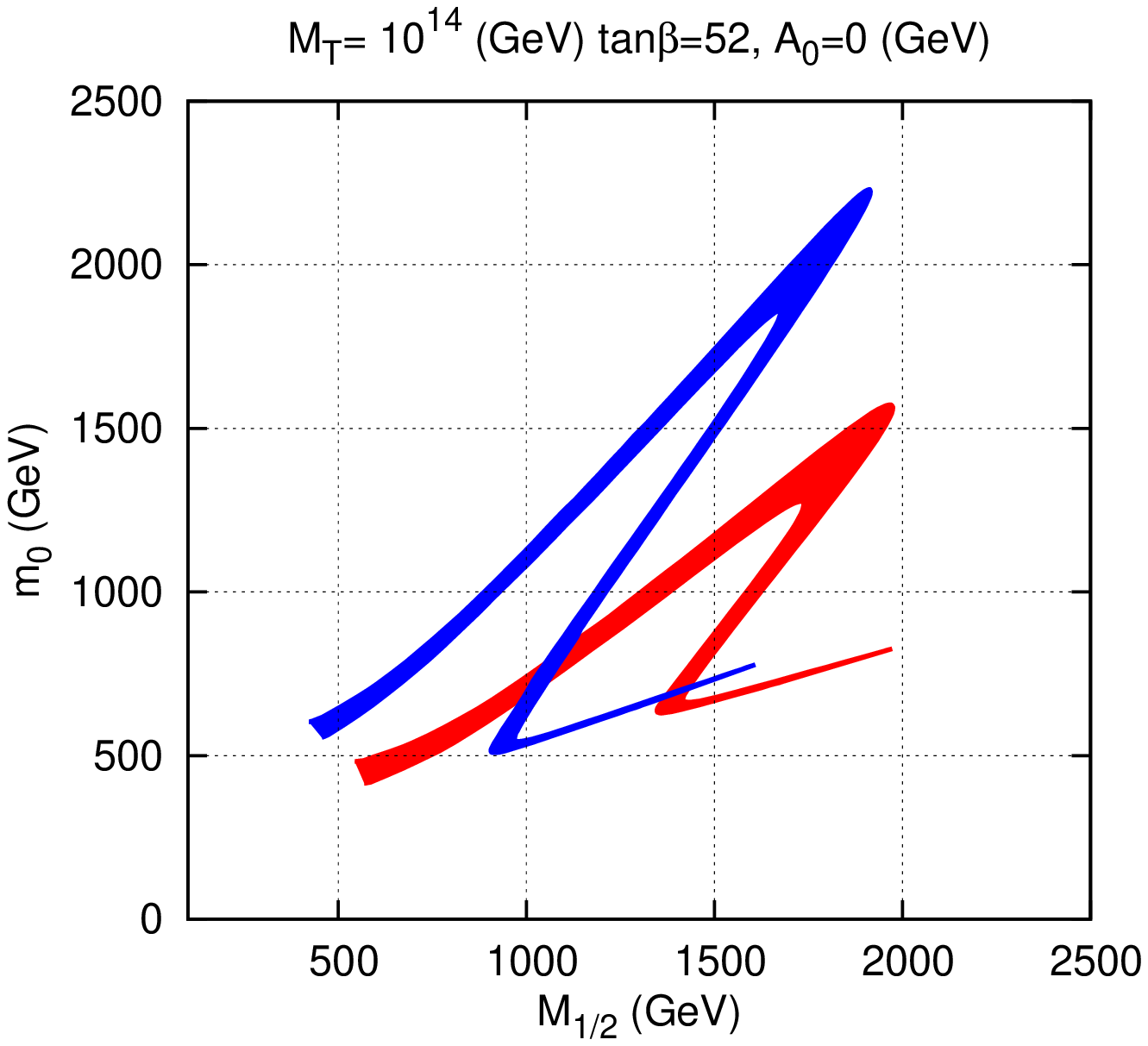}&
  \includegraphics[width=0.45\linewidth]{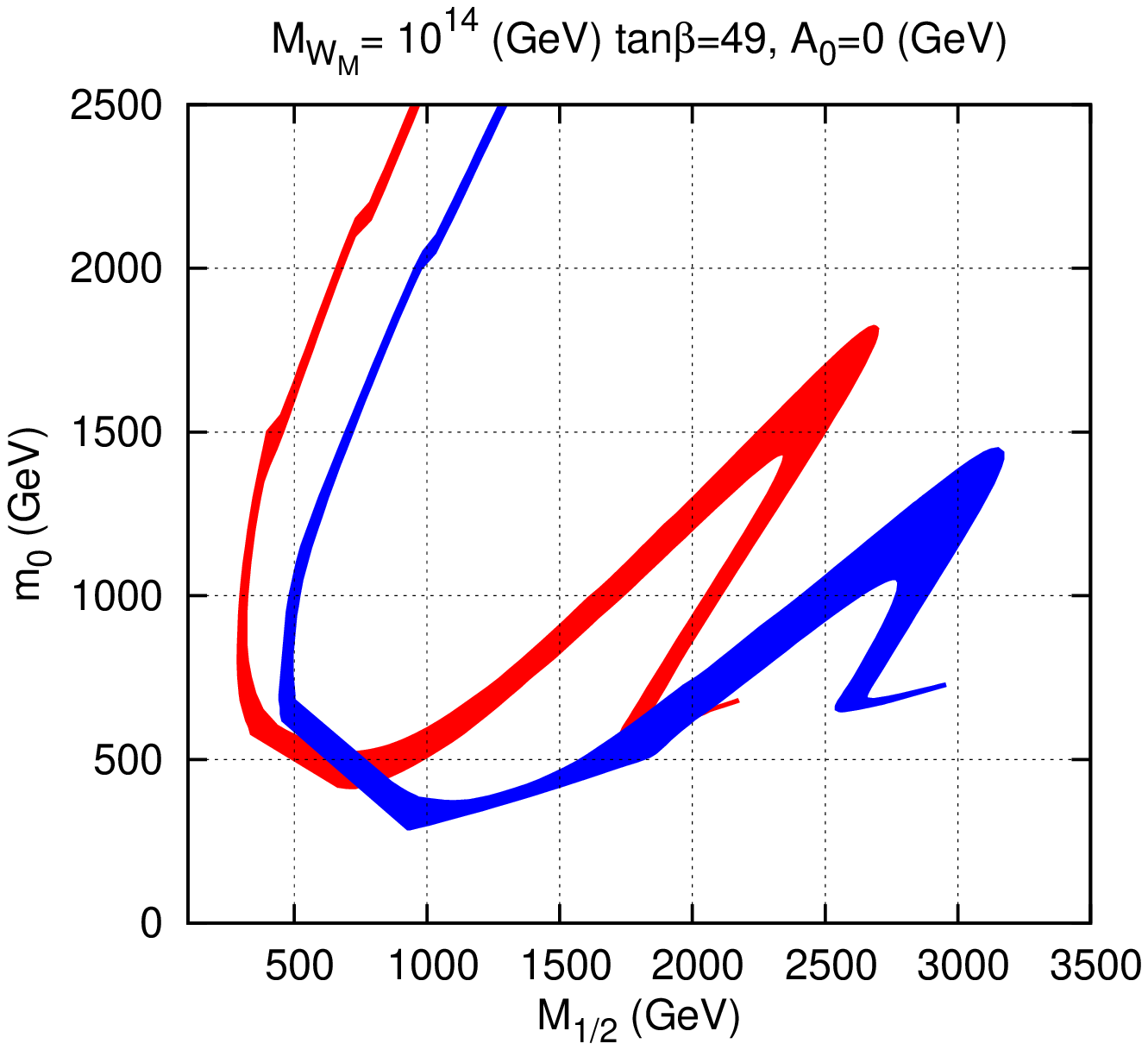}
\end{tabular}
   \caption{Comparison between using 1-loop (red) or 2-loop
     (blue) RGEs on the dark matter allowed region for type-II (left
     panel) and type-III (right panel).  
     The parameters are: $A_0=0$, $\mu>0$
     and $M_{\rm Seesaw}=10^{14}$ GeV, $m_{top}=171.2$ GeV and
     $\tan\beta=52$ for type-II and $\tan\beta=49$ for type-III. }
   \label{fig:4}
\end{figure}

In the case of large $\tan\beta$ an additional region, usually called 
the Higgs funnel, opens up. This region is characterized by $M_A \simeq 2
m_{\tilde \chi^0_1}$.  Also here the regions gets shifted compared to
usual mSUGRA scenario. However, this region is very sensitive to
higher order corrections and therefore it is quite important to use 
full 2-loop RGEs as can be seen in Fig.~\ref{fig:4}. We have again fixed 
$A_0$ = 0, $\mu > 0$, $m_{top}=171.2$ GeV and the seesaw scale to
$10^{14}$ GeV, with a degenerate spectrum in case of the type-III model.
The main reason for the observed and rather surprisingly large differences 
between the different calculations is that the 2-loop
contributions decrease the neutralino mass compared to the 1-loop case
while at the same time increasing $M_A$. For example, in case of
seesaw II and for fixed values of $m_0 = M_{1/2}$ = 1500 GeV we get in
case of 1-loop RGEs $m_{\tilde\chi^0_1} = 560$ GeV, $M_A =
1090$~GeV and in case of 2-loop RGEs $m_{\tilde\chi^0_1} = 498$
GeV, $M_A = 1100$~GeV.  For completeness we note that this region
is also very sensitive to input values for $m_t$ and $m_b$
\cite{Esteves:2009qr}.

%\clearpage

\section{Conclusions and outlook}
\label{sec:cncl}

To summarize, we have investigated in detail a supersymmetric version
of a seesaw model of type-III and compared it to seesaw models of
type-I and type-II. In case of type-II and type-III models we have
embedded the $SU(2)$ triplets in the corresponding $SU(5)$
representations to maintain gauge coupling unification, e.g.~{\bf
15}-plets in case of type-II and {\bf 24}-plets in case of type-III
models. For definiteness we have assumed mSUGRA boundary conditions for
the soft SUSY breaking parameters.

The additional heavy charged states lead to changes in the
beta-functions and, thus, also in the running of the SUSY mass
parameters.  We have calculated the soft masses as a function of the
seesaw parameters. As discussed in some detail, there are certain
combinations of soft masses, which are approximately constants over
large regions of mSUGRA space. These ``invariants'' contain indirect
information about the seesaw scale assuming the type of seesaw model.
In certain parts of the parameter space, e.g.~for low seesaw scales,
one might even be able to exclude certain seesaw models by combining
mass measurements at the LHC with the mSUGRA paradigm. We note, that
using 2-loop RGEs will be crucial to obtain reliable results.

The changes in the spectrum leads obviously to changes in the
phenomenology.  We have calculated lepton flavour violating
observables, such as $Br(l_i\to l_j+\gamma)$. We find that for fixed
(degenerate) seesaw scale these branching ratios are in general
largest for type-III models followed by type-II and type-I. This is a
consequence of the fact that for a given set of mSUGRA parameters the
spectrum in type-III is lighter than for type-II models which is again
lighter than in type-I models. However, the difference in the predictions of type-II and type-III
is somewhat smaller than expected from these considerations because
in type-II models the flavour violating entries
are larger compared to the case of type-III models.

We also investigated the predictions for the relic density $\Omega
h^2$ in the type-III model and compared them with the other models. We
find the usual four regions in the mSUGRA parameter space but of
course they are shifted due to the changes in the spectrum.  It has been 
found that in particular in case of the Higgs-funnel the use of 2-loop
RGEs is crucial to identify the correct allowed region. Last but not
least we note, that for low seesaw scales the co-annihilation region
vanishes for both, the type-II and the type-III models, as the
required mass difference between the lightest neutralino and the stau
cannot be obtained anymore.

\section*{Acknowledgments}

W.P. thanks
 IFIC/C.S.I.C. for hospitality during an extended stay.
This work was supported by the Spanish MICINN under grants
FPA2008-00319/FPA, by the MULTIDARK Consolider CAD2009-00064, by
Prometeo/2009/091, by the EU grant UNILHC PITN-GA-2009-237920 
and FPA2008-04002-E/PORTU. 
The work of J.~N.~E.  has been supported by {\it Funda\c c\~ao para a
Ci\^encia e a Tecnologia} through the fellowship SFRH/BD/29642/2006.
J.~N.~E. and J.~C.~R. also acknowledge the financial support from
 {\it Funda\c{c}\~ao para a Ci\^encia e a Tecnologia} grants CFTP-FCT
UNIT~777 and CERN/FP/109305/2009.
W.P. is partially  supported by the German
Ministry of Education and Research (BMBF) under contract 05HT6WWA
 and by the Alexander von Humboldt Foundation. F.S.\ has been supported by the
DFG research training group GRK1147.

\appendix

\section{RGEs for the seesaw type-II and seesaw type-III models at 2-loop}
\label{app:RGEs}

In the appendix we collect the beta coefficients for the gauge couplings
as well as anomalous dimensions of the superfields which are the
ingredients to calculate the 2-loop
RGEs for both, the seesaw type-II and type-III, models using the 
procedure given in \cite{Jack:1997eh} based on the spurion formalism 
\cite{Yamada:1994id}. The complete set of RGEs for both models at 2-loop is also given online \cite{SarahWeb}.
In the following we briefly summarize the basic
ideas of this calculation for completeness.

For a  general $N=1$ supersymmetric gauge theory with superpotential  
\begin{equation}
 W (\phi) = \frac{1}{2}{\mu}^{ij}\phi_i\phi_j + \frac{1}{6}Y^{ijk}
\phi_i\phi_j\phi_k
\end{equation}
the  soft SUSY-breaking scalar terms are given by
\begin{equation}
V_{\hbox{soft}} = \left(\frac{1}{2}b^{ij}\phi_i\phi_j
+ \frac{1}{6}h^{ijk}\phi_i\phi_j\phi_k +\hbox{c.c.}\right)
+(m^2)^i{}_j\phi_i\phi_j^*.
\end{equation}

The $\beta$-functions for the superpotential parameters can be obtained by using superfield technique \cite{West:1984dg,Jones:1984cx}. The obtained results are \cite{Martin:1993zk}
\begin{eqnarray}
 \beta_Y^{ijk} &= & Y^{p(ij} {\gamma_p}^{k)} \thickspace, \\
 \beta_{\mu}^{ij} &= & \mu^{p(i} {\gamma_p}^{j)} \thickspace.
\end{eqnarray}

The  exact results for the soft $\beta$-functions
are given by \cite{Jack:1997eh}:
\begin{eqnarray}
\label{eq:betaM}
\beta_M &=& 2{\cal O} \left[\frac{\beta_g}{g}\right], \\
\beta_{h}^{ijk} &=& h{}^{l(jk}\gamma^{i)}{}_l -
2Y^{l(jk}\gamma_1{}^{i)}{}_l, \cr
\beta_{b}^{ij} &=&   
b{}^{l(i}\gamma^{j)}{}_l-2\mu{}^{l(i}\gamma_1{}^{j)}{}_l,\cr
\left(\beta_{m^2}\right){}^i{}_j &=& \Delta\gamma^i{}_j
\label{eq:betam2}
\end{eqnarray}
where $\gamma$ is the matter multiplet anomalous dimension, $\beta_g$ the
beta function for the gauge coupling $g$; the $(..)$ in the
superscripts denote symmetrisation and  
\begin{eqnarray}
{\cal O}  &=& Mg^2\frac{\partial}{\partial g^2}-h^{lmn}
\frac{\partial}{\partial Y^{lmn}} \thickspace, \\
(\gamma_1)^i{}_j  &=& {\cal O}\gamma^i{}_j, \\
\Delta &=& 2{\cal O} {\cal O}^* +2MM^* g^2{\partial\over{\partial g^2}}
+\left[{\tilde Y}^{lmn}{\partial\over{\partial Y^{lmn}}} + \hbox{c.c.}\right]
+X{\partial\over{\partial g}} \thickspace.
\end{eqnarray}
Here $M$ is the gaugino mass and  
${\tilde Y}^{ijk} = (m^2)^i{}_lY^{jkl} +  (m^2)^j{}_lY^{ikl} + (m^2)^k{}_lY^{ijl}.$
Equations.~(\ref{eq:betaM})--(\ref{eq:betam2}) hold
 in a class of renormalisation schemes that includes 
the DRED$'$-one \cite{Jack:1994rk}. We take the known contributions of $X$
from \cite{Jack:1998iy}:
\begin{eqnarray}
X^{\mathrm{DRED}'(1)}&=&-2g^3S, \\
X^{\mathrm{DRED}'(2)}&=& (2r)^{-1}g^3 \mathrm{tr} [ W C(R)]
-4g^5C(G)S-2g^5C(G)QMM^*,\end{eqnarray}
where
\begin{eqnarray}
S &=&  r^{-1} \mathrm{tr}[m^2C(R)] -MM^* C(G),  \\
W^j{}_i&=&{1\over2}Y_{ipq}Y^{pqn}(m^2)^j{}_n+{1\over2}Y^{jpq}Y_{pqn}(m^2)^n{}_i
+2Y_{ipq}Y^{jpr}(m^2)^q{}_r 
+h_{ipq}h^{jpq}-8g^2MM^*C(R)^j{}_i,
\nonumber \\
\end{eqnarray}
$C(R),C(G)$ being the quadratic Casimirs for the matter and adjoint 
representations, respectively,
 $Q = T(R) - 3C(G)$, and $rT(R) = \mathrm{tr} \left[C(R)\right]$, $r$ being 
the number of group generators. 

In the following subsections we give the anomalous dimensions and beta functions
for the seesaw models of type-II and type-III so that with the help of the above
equations one can calculate all RGEs at the 2-loop level.

\subsection{The anomalous dimensions for seesaw type-II}
\label{app:AnoII}
Here and in the subsequent sections ${\bf 1}$ denotes the 3$\times$3 unit matrix. \(N_X\) is the number of generations of heavy field \(X\). Furthermore, we define
\begin{equation}
 \tilde{N}_X = N_X + N_{\bar{X}} \thickspace.
\end{equation}

{\allowdisplaybreaks \begin{align} 
\gamma_{\hat{q}}^{(1)}  = \, &
-\frac{1}{30} \Big(45 g_{2}^{2}  + 80 g_{3}^{2}  + g_{1}^{2}\Big){\bf 1}  + {Y_{d}^{\dagger}  Y_d} + {Y_{u}^{\dagger}  Y_u}\\ 
\gamma_{\hat{q}}^{(2)}  = \, &
+\frac{4}{5} g_{1}^{2} {Y_{u}^{\dagger}  Y_u} -3 |\lambda_2|^2 {Y_{u}^{\dagger}  Y_u} -2 {Y_{d}^{\dagger}  Y_d  Y_{d}^{\dagger}  Y_d} -4 {Y_{d}^{\dagger}  Y_s  Y_s^*  Y_d} -2 {Y_{d}^{\dagger}  Y_z  Y_{z}^{\dagger}  Y_d} -2 {Y_{u}^{\dagger}  Y_u  Y_{u}^{\dagger}  Y_u} \nonumber \\ 
 &+{\bf 1} \Big[199 g_{1}^{4} +90 g_{1}^{2} g_{2}^{2} +3375 g_{2}^{4} +160(g_{1}^{2} g_{3}^{2} +5(4 g_{2}^{2} g_{3}^{2} - g_{3}^{4})) +48 \Big(125 g_{3}^{4}  + g_{1}^{4}\Big)\tilde{N}_S  \nonumber \\ 
 &+\left(54 g_{1}^{4}+2700 g_{2}^{4}\right) \tilde{N}_{T}+ \left(3 g_{1}^{4}+2025 g_{2}^{4}+2400 g_3^4 \right) \tilde{N}_{Z} \Big] \frac{1}{900} \nonumber \\ 
& +{Y_{d}^{\dagger}  Y_d} \Big[-3 |\lambda_1|^2  -3 \mbox{Tr}\Big({Y_d  Y_{d}^{\dagger}}\Big)  + \frac{2}{5} g_{1}^{2} - \mbox{Tr}\Big({Y_e  Y_{e}^{\dagger}}\Big) \Big]-3 {Y_{u}^{\dagger}  Y_u} \mbox{Tr}\Big({Y_u  Y_{u}^{\dagger}}\Big) \\ 
\gamma_{\hat{l}}^{(1)}  = \, &
3 \Big({Y_{z}^{\dagger}  Y_z} + {Y_t^*  Y_t}\Big) -\frac{3}{10} \Big(5 g_{2}^{2}  + g_{1}^{2}\Big){\bf 1}  + {Y_{e}^{\dagger}  Y_e}\\ 
\gamma_{\hat{l}}^{(2)}  = \, &
-\frac{2}{5} g_{1}^{2} {Y_{z}^{\dagger}  Y_z} +16 g_{3}^{2} {Y_{z}^{\dagger}  Y_z} +\frac{18}{5} g_{1}^{2} {Y_t^*  Y_t} +12 g_{2}^{2} {Y_t^*  Y_t} -3 |\lambda_1|^2 {Y_t^*  Y_t} -2 {Y_{e}^{\dagger}  Y_e  Y_{e}^{\dagger}  Y_e} \nonumber \\ 
 &-6 {Y_{z}^{\dagger}  Y_d  Y_{d}^{\dagger}  Y_z} -12 {Y_{z}^{\dagger}  Y_s  Y_s^*  Y_z} -6 {Y_{z}^{\dagger}  Y_z  Y_{z}^{\dagger}  Y_z} -9 {Y_t^*  Y_t  Y_t^*  Y_t} -3 {Y_t^*  Y_{e}^{T}  Y_e^*  Y_t} \nonumber \\ 
 & -9 {Y_t^*  Y_{z}^{T}  Y_z^*  Y_t} +\frac{3}{100} {\bf 1} \Big[69 g_{1}^{4} +30 g_{1}^{2} g_{2}^{2} +125 g_{2}^{4} +16 g_{1}^{4} \tilde{N}_S +\left(18 g_{1}^{4}  +100 g_{2}^{4}  \right)\tilde{N}_T \nonumber \\ 
 &+\left(g_{1}^{4}  +75 g_{2}^{4} \right)\tilde{N}_Z \Big]+{Y_{e}^{\dagger}  Y_e} \Big[-3 |\lambda_1|^2  -3 \mbox{Tr}\Big({Y_d  Y_{d}^{\dagger}}\Big)  + \frac{6}{5} g_{1}^{2}  - \mbox{Tr}\Big({Y_e  Y_{e}^{\dagger}}\Big) \Big]\nonumber \\ 
 &-3 {Y_t^*  Y_t} \mbox{Tr}\Big({Y_t  Y_t^*}\Big) -3 {Y_{z}^{\dagger}  Y_z} \mbox{Tr}\Big({Y_z  Y_{z}^{\dagger}}\Big) \\ 
\gamma_{\hat{H}_d}^{(1)}  = \, &
3 |\lambda_1|^2  + 3 \mbox{Tr}\Big({Y_d  Y_{d}^{\dagger}}\Big)  -\frac{3}{10} g_{1}^{2}  -\frac{3}{2} g_{2}^{2}  + \mbox{Tr}\Big({Y_e  Y_{e}^{\dagger}}\Big)\\ 
\gamma_{\hat{H}_d}^{(2)}  = \, &
-12 |\lambda_{1}^{2}|^4+\frac{3}{5} |\lambda_1|^2 \Big[-15 \mbox{Tr}\Big({Y_d  Y_{d}^{\dagger}}\Big)  + 20 g_{2}^{2}  -5 \mbox{Tr}\Big({Y_e  Y_{e}^{\dagger}}\Big)  -5 \mbox{Tr}\Big({Y_t  Y_t^*}\Big)  + 6 g_{1}^{2} \Big]\nonumber \\ 
 &+\frac{1}{100} \Big[207 g_{1}^{4} +90 g_{1}^{2} g_{2}^{2} +375 g_{2}^{4} +48 g_{1}^{4}\tilde{N}_S +\left(54 g_{1}^{4} +300 g_{2}^{4} \right)\tilde{N}_T \nonumber \\ 
 &+ (3 g_{1}^{4} +225 g_{2}^{4})\tilde{N}_Z -40 g_{1}^{2} \mbox{Tr}\Big({Y_d  Y_{d}^{\dagger}}\Big)\Big] +16 g_{3}^{2} \mbox{Tr}\Big({Y_d  Y_{d}^{\dagger}}\Big)  -9 \mbox{Tr}\Big({Y_d  Y_{d}^{\dagger}  Y_d  Y_{d}^{\dagger}}\Big) \nonumber \\ 
 &-12 \mbox{Tr}\Big({Y_d  Y_{d}^{\dagger}  Y_s  Y_s^*}\Big) -6 \mbox{Tr}\Big({Y_d  Y_{d}^{\dagger}  Y_z  Y_{z}^{\dagger}}\Big) -3 \mbox{Tr}\Big({Y_d  Y_{u}^{\dagger}  Y_u  Y_{d}^{\dagger}}\Big) -3 \mbox{Tr}\Big({Y_e  Y_{e}^{\dagger}  Y_e  Y_{e}^{\dagger}} \nonumber \\ 
 &-3 \mbox{Tr}\Big({Y_e  Y_{z}^{\dagger}  Y_z  Y_{e}^{\dagger}}\Big) -3 \mbox{Tr}\Big({Y_e  Y_t^*  Y_t  Y_{e}^{\dagger}}\Big)+1.2 g_{1}^{2} \mbox{Tr}\Big({Y_e  Y_{e}^{\dagger}}\Big) \Big)\\ 
\gamma_{\hat{H}_u}^{(1)}  = \, &
3 |\lambda_2|^2  -\frac{3}{10} \Big(-10 \mbox{Tr}\Big({Y_u  Y_{u}^{\dagger}}\Big)  + 5 g_{2}^{2}  + g_{1}^{2}\Big)\\ 
\gamma_{\hat{H}_u}^{(2)}  = \, &
\frac{1}{100} \Big[207 g_{1}^{4} +90 g_{1}^{2} g_{2}^{2} +375 g_{2}^{4} -1200 |\lambda_{2}|^4 +48 g_{1}^{4}\tilde{N}_S + \left(54 g_{1}^{4}+300 g_{2}^{4}\right)\tilde{N}_T  \nonumber \\ 
 &+\left(3 g_{1}^{4} +225 g_{2}^{4}\right)\tilde{N}_Z +60 |\lambda_2|^2 \Big(-15 \mbox{Tr}\Big({Y_u  Y_{u}^{\dagger}}\Big)  + 20 g_{2}^{2}  + 6 g_{1}^{2} \Big)+80 g_{1}^{2} \mbox{Tr}\Big({Y_u  Y_{u}^{\dagger}}\Big) \Big] \nonumber \\ 
 &+16 g_{3}^{2} \mbox{Tr}\Big({Y_u  Y_{u}^{\dagger}}\Big) -3 \mbox{Tr}\Big({Y_d  Y_{u}^{\dagger}  Y_u  Y_{d}^{\dagger}}\Big) -9\mbox{Tr}\Big({Y_u  Y_{u}^{\dagger}  Y_u  Y_{u}^{\dagger}}\Big) \\ 
\gamma_{\hat{d}}^{(1)}  = \, &
2 \Big(2 {Y_s^*  Y_s}  + {Y_d^*  Y_{d}^{T}} + {Y_z^*  Y_{z}^{T}}\Big) -\frac{2}{15} \Big(20 g_{3}^{2}  + g_{1}^{2}\Big){\bf 1} \\ 
\gamma_{\hat{d}}^{(2)}  = \, &
+\frac{32}{15} g_{1}^{2} {Y_s^*  Y_s} +\frac{80}{3} g_{3}^{2} {Y_s^*  Y_s} +\frac{2}{5} g_{1}^{2} {Y_z^*  Y_{z}^{T}} +6 g_{2}^{2} {Y_z^*  Y_{z}^{T}} -2 {Y_d^*  Y_{d}^{T}  Y_d^*  Y_{d}^{T}} -2 {Y_d^*  Y_{u}^{T}  Y_u^*  Y_{d}^{T}} \nonumber \\ 
 &-8 {Y_s^*  Y_d  Y_{d}^{\dagger}  Y_s} -16 {Y_s^*  Y_s  Y_s^*  Y_s} -8 {Y_s^*  Y_z  Y_{z}^{\dagger}  Y_s} -6 {Y_z^*  Y_t  Y_t^*  Y_{z}^{T}} -2 {Y_z^*  Y_{e}^{T}  Y_e^*  Y_{z}^{T}} -6 {Y_z^*  Y_{z}^{T}  Y_z^*  Y_{z}^{T}} \nonumber \\ 
 &+\frac{1}{225} {\bf 1} \Big[202 g_{1}^{4} +160 g_{1}^{2} g_{3}^{2} -200 g_{3}^{4} +12 \Big(125 g_{3}^{4}  + 4 g_{1}^{4} \Big)\tilde{N}_S +54 g_{1}^{4}\tilde{N}_T +\left(3 g_{1}^{4}  +600 g_{3}^{4}\right)\tilde{N}_Z \Big]\nonumber \\ 
 &-2 {Y_z^*  Y_{z}^{T}} \mbox{Tr}\Big({Y_z  Y_{z}^{\dagger}}\Big)-4 {Y_s^*  Y_s} \mbox{Tr}\Big({Y_s  Y_s^*}\Big)+{Y_d^*  Y_{d}^{T}} \Big[-2 \mbox{Tr}\Big({Y_e  Y_{e}^{\dagger}}\Big)  + 6 g_{2}^{2}  -6 |\lambda_1|^2  \nonumber \\ &-6 \mbox{Tr}\Big({Y_d  Y_{d}^{\dagger}}\Big)   + \frac{2}{5} g_{1}^{2} \Big]  \\ 
\gamma_{\hat{u}}^{(1)}  = \, &
2 {Y_u^*  Y_{u}^{T}}  -\frac{8}{15} \Big(5 g_{3}^{2}  + g_{1}^{2}\Big){\bf 1} \\ 
\gamma_{\hat{u}}^{(2)}  = \, &
\frac{2}{225} \Big[2 {\bf 1} \Big(214 g_{1}^{4} +160 g_{1}^{2} g_{3}^{2} -50 g_{3}^{4} +\Big(375 g_{3}^{4}  + 48 g_{1}^{4} \Big) \tilde{N}_S +54 g_{1}^{4} \tilde{N}_T + \left(3 g_{1}^{4} +150 g_{3}^{4}\right)\tilde{N}_Z \Big)\nonumber \\ 
 &-45 \Big\{5 \Big({Y_u^*  Y_{d}^{T}  Y_d^*  Y_{u}^{T}} + {Y_u^*  Y_{u}^{T}  Y_u^*  Y_{u}^{T}}\Big) + {Y_u^*  Y_{u}^{T}} \Big(-15 g_{2}^{2}  + 15 |\lambda_2|^2  + 15 \mbox{Tr}\Big({Y_u  Y_{u}^{\dagger}}\Big)  + g_{1}^{2}\Big)\Big\}\Big]\\ 
\gamma_{\hat{e}}^{(1)}  = \, &
2 {Y_e^*  Y_{e}^{T}}  -\frac{6}{5} g_{1}^{2} {\bf 1} \\ 
\gamma_{\hat{e}}^{(2)}  = \, &
\frac{1}{25} \Big[3 g_{1}^{4} {\bf 1} \Big(16\tilde{N}_S  + 18\tilde{N}_T  + 78 + \tilde{N}_Z\Big)-10 \Big\{5 \Big(3 {Y_e^*  Y_t  Y_t^*  Y_{e}^{T}}  + 3 {Y_e^*  Y_{z}^{T}  Y_z^*  Y_{e}^{T}} \nonumber \\ 
 & + {Y_e^*  Y_{e}^{T}  Y_e^*  Y_{e}^{T}}\Big)+{Y_e^*  Y_{e}^{T}} \Big(-15 g_{2}^{2}  + 15 |\lambda_1|^2  + 15 \mbox{Tr}\Big({Y_d  Y_{d}^{\dagger}}\Big)  + 3 g_{1}^{2}  + 5 \mbox{Tr}\Big({Y_e  Y_{e}^{\dagger}}\Big) \Big)\Big\}\Big]\\ 
\gamma_{\hat{T}}^{(1)}  = \, &
-4 g_{2}^{2}  -\frac{6}{5} g_{1}^{2}  + |\lambda_1|^2 + \mbox{Tr}\Big({Y_t  Y_t^*}\Big)\\ 
\gamma_{\hat{T}}^{(2)}  = \, &
\frac{1}{25} \Big[234 g_{1}^{4} +240 g_{1}^{2} g_{2}^{2} +500 g_{2}^{4} -150 |\lambda_{1}|^4  +48 g_{1}^{4} \tilde{N}_S  +\left(54 g_{1}^{4}+200 g_{2}^{4}\right)\tilde{N}_T \nonumber \\ 
 &+ \left(3 g_{1}^{4}  +150 g_{2}^{4}\right)\tilde{N}_Z -5 |\lambda_1|^2 \Big(10 \mbox{Tr}\Big({Y_e  Y_{e}^{\dagger}}\Big)  + 30 \mbox{Tr}\Big({Y_d  Y_{d}^{\dagger}}\Big)  + 3 g_{1}^{2}  + 5 g_{2}^{2} \Big)\nonumber \\ 
 &-15 g_{1}^{2} \mbox{Tr}\Big({Y_t  Y_t^*}\Big) -25 g_{2}^{2} \mbox{Tr}\Big({Y_t  Y_t^*}\Big) -50 \mbox{Tr}\Big({Y_e  Y_t^*  Y_t  Y_{e}^{\dagger}}\Big) -150 \mbox{Tr}\Big({Y_t  Y_{z}^{\dagger}  Y_z  Y_t^*}\Big) \nonumber \\
& -150 \mbox{Tr}\Big({Y_t  Y_t^*  Y_t  Y_t^*}\Big) \Big]\\ 
\gamma_{\hat{\bar{T}}}^{(1)}  = \, &
-4 g_{2}^{2}  -\frac{6}{5} g_{1}^{2}  + |\lambda_2|^2\\ 
\gamma_{\hat{\bar{T}}}^{(2)}  = \, &
\frac{1}{25} \Big[234 g_{1}^{4} +240 g_{1}^{2} g_{2}^{2} +500 g_{2}^{4} -150 |\lambda_{2}|^4 +48 g_{1}^{4} \tilde{N}_S +\left(54 g_{1}^{4} +200 g_{2}^{4}\right)\tilde{N}_T \nonumber \\ 
 &+\left(3 g_{1}^{4} +150 g_{2}^{4}\right)\tilde{N}_Z  -5 |\lambda_2|^2 \Big(30 \mbox{Tr}\Big({Y_u  Y_{u}^{\dagger}}\Big)  + 3 g_{1}^{2}  + 5 g_{2}^{2} \Big)\Big]\\ 
\gamma_{\hat{S}}^{(1)}  = \, &
-\frac{4}{15} \Big(25 g_{3}^{2}  + 2 g_{1}^{2} \Big) + \mbox{Tr}\Big({Y_s  Y_s^*}\Big)\\ 
\gamma_{\hat{S}}^{(2)}  = \, &
\frac{2}{225} \Big[3 \Big(32 g_{1}^{4}  + 625 g_{3}^{4} \Big)\tilde{N}_S +2 \Big\{214 g_{1}^{4} +400 g_{1}^{2} g_{3}^{2} +1375 g_{3}^{4} +54 g_{1}^{4} \tilde{N}_T \nonumber \\ 
 & + \left(3 g_{1}^{4} +375 g_{3}^{4}\right)\tilde{N}_Z -15 g_{1}^{2} \mbox{Tr}\Big({Y_s  Y_s^*}\Big) -75 g_{3}^{2} \mbox{Tr}\Big({Y_s  Y_s^*}\Big) -225 \mbox{Tr}\Big({Y_d  Y_{d}^{\dagger}  Y_s  Y_s^*}\Big) \nonumber \\ 
 &-450 \mbox{Tr}\Big({Y_s  Y_s^*  Y_s  Y_s^*}\Big) -225 \mbox{Tr}\Big({Y_s  Y_s^*  Y_z  Y_{z}^{\dagger}}\Big) \Big\}\Big]\\ 
\gamma_{\hat{\bar{S}}}^{(1)}  = \, &
-\frac{4}{15} \Big(25 g_{3}^{2}  + 2 g_{1}^{2} \Big)\\ 
\gamma_{\hat{\bar{S}}}^{(2)}  = \, &
\frac{2}{225} \Big[428 g_{1}^{4} +800 g_{1}^{2} g_{3}^{2} +2750 g_{3}^{4} +3 \Big(32 g_{1}^{4}  + 625 g_{3}^{4} \Big)\tilde{N}_S+108 g_{1}^{4}\tilde{N}_T + \left(6 g_{1}^{4}+750 g_{3}^{4}\right)\tilde{N}_Z \Big]\\ 
\gamma_{\hat{Z}}^{(1)}  = \, &
\frac{1}{30} \Big(30 \mbox{Tr}\Big({Y_z  Y_{z}^{\dagger}}\Big)  -45 g_{2}^{2}  -80 g_{3}^{2}  - g_{1}^{2} \Big)\\ 
\gamma_{\hat{Z}}^{(2)}  = \, &
+\frac{199}{900} g_{1}^{4} +\frac{1}{10} g_{1}^{2} g_{2}^{2} +\frac{15}{4} g_{2}^{4} +\frac{8}{45} g_{1}^{2} g_{3}^{2} +8 g_{2}^{2} g_{3}^{2} -\frac{8}{9} g_{3}^{4} +\frac{4}{75} \Big(125 g_{3}^{4}  + g_{1}^{4}\Big)\tilde{N}_S \nonumber \\ 
 & +\left(\frac{3}{50} g_{1}^{4} +3 g_{2}^{4}\right)\tilde{N}_T +\left(\frac{1}{300} g_{1}^{4} +\frac{9}{4} g_{2}^{4}+\frac{8}{3} g_{3}^{4}\right)\tilde{N}_Z +\frac{2}{5} g_{1}^{2} \mbox{Tr}\Big({Y_z  Y_{z}^{\dagger}}\Big) -2 \mbox{Tr}\Big({Y_d  Y_{d}^{\dagger}  Y_z  Y_{z}^{\dagger}}\Big)\nonumber \\ 
 & - \mbox{Tr}\Big({Y_e  Y_{z}^{\dagger}  Y_z  Y_{e}^{\dagger}}\Big) -4 \mbox{Tr}\Big({Y_s  Y_s^*  Y_z  Y_{z}^{\dagger}}\Big) -3 \mbox{Tr}\Big({Y_t  Y_{z}^{\dagger}  Y_z  Y_t^*}\Big)-5 \mbox{Tr}\Big({Y_z  Y_{z}^{\dagger}  Y_z  Y_{z}^{\dagger}}\Big) \\ 
\gamma_{\hat{\bar{Z}}}^{(1)}  = \, &
\frac{1}{30} \Big(-45 g_{2}^{2}  -80 g_{3}^{2}  - g_{1}^{2} \Big)\\ 
\gamma_{\hat{\bar{Z}}}^{(2)}  = \, &
\frac{1}{900} \Big[199 g_{1}^{4} +90 g_{1}^{2} g_{2}^{2} +3375 g_{2}^{4} +160(g_{1}^{2} g_{3}^{2} +20 g_{2}^{2} g_{3}^{2} -5 g_{3}^{4}) +48 \Big(125 g_{3}^{4}  + g_{1}^{4}\Big)\tilde{N}_S \nonumber \\ 
 &+\left(54 g_{1}^{4} +2700 g_{2}^{4}\right)\tilde{N}_T +\left(3 g_{1}^{4}+2025 g_{2}^{4} +2400 g_{3}^{4}\right)\tilde{N}_Z \Big]
\end{align} } 
\subsection{Beta coefficients for the seesaw type-II at 2-loop level}
\label{app:BetaII}
{\allowdisplaybreaks  \begin{align} 
\beta_{g_1}^{(1)}  = \, &
\frac{1}{10} g_{1}^{3} \Big(16 \tilde{N}_S  + 18 \tilde{N}_T  + 66 + \tilde{N}_Z\Big)\\ 
\beta_{g_1}^{(2)}  = \, &
\frac{1}{150} g_{1}^{3} \Big[1194 g_{1}^{2} +810 g_{2}^{2} +2640 g_{3}^{2} -810(|\lambda_1|^2 +|\lambda_2|^2) +\left(256 g_{1}^{2} +3200 g_{3}^{2}\right)\tilde{N}_S  \nonumber \\ 
 &+\left(648 g_{1}^{2} +2160 g_{2}^{2}\right)\tilde{N}_T +\left(g_{1}^{2}+45 g_{2}^{2} +80 g_{3}^{2}\right)\tilde{N}_Z -420 \mbox{Tr}\Big({Y_d  Y_{d}^{\dagger}}\Big) \nonumber \\ 
 &-540 \mbox{Tr}\Big({Y_e  Y_{e}^{\dagger}}\Big) -720 \mbox{Tr}\Big({Y_s  Y_s^*}\Big) -810 \mbox{Tr}\Big({Y_t  Y_t^*}\Big) -780 \mbox{Tr}\Big({Y_u  Y_{u}^{\dagger}}\Big) -420 \mbox{Tr}\Big({Y_z  Y_{z}^{\dagger}}\Big) \Big]\\ 
\beta_{g_2}^{(1)}  = \, &
\frac{1}{2} g_{2}^{3} \Big(3 \tilde{N}_Z  + 4 \tilde{N}_T  + 2\Big)\\ 
\beta_{g_2}^{(2)}  = \, &
\frac{1}{10} g_{2}^{3} \Big[18 g_{1}^{2} +250 g_{2}^{2} +240 g_{3}^{2} -70 |\lambda_1|^2 -70 |\lambda_2|^2 +\left(48 g_{1}^{2} +240 g_{2}^{2} \right)\tilde{N}_T \nonumber \\ 
 &+\left(g_{1}^{2}  +105 g_{2}^{2} +80 g_{3}^{2}\right)\tilde{N}_Z -60 \mbox{Tr}\Big({Y_d  Y_{d}^{\dagger}}\Big) -20 \mbox{Tr}\Big({Y_e  Y_{e}^{\dagger}}\Big) -70 \mbox{Tr}\Big({Y_t  Y_t^*}\Big)\nonumber \\ 
 & -60 \mbox{Tr}\Big({Y_u  Y_{u}^{\dagger}}\Big) -60 \mbox{Tr}\Big({Y_z  Y_{z}^{\dagger}}\Big) \Big]\\ 
\beta_{g_3}^{(1)}  = \, &
\frac{1}{2} g_{3}^{3} \Big(2 \Big(-3 + \tilde{N}_Z\Big) + 5\tilde{N}_S \Big)\\ 
\beta_{g_3}^{(2)}  = \, &
\frac{1}{15} g_{3}^{3} \Big[33 g_{1}^{2} +135 g_{2}^{2} +210 g_{3}^{2} +5 \Big(145 g_{3}^{2}  + 8 g_{1}^{2} \Big)\tilde{N}_S-135 \mbox{Tr}\Big({Y_s  Y_s^*}\Big) \nonumber \\ & +\left(g_{1}^{2} +45 g_{2}^{2} +170 g_{3}^{2}\right)\tilde{N}_Z -60 \mbox{Tr}\Big({Y_d  Y_{d}^{\dagger}}\Big) -60 \mbox{Tr}\Big({Y_u  Y_{u}^{\dagger}}\Big) -60 \mbox{Tr}\Big({Y_z  Y_{z}^{\dagger}}\Big) \Big]
\end{align}} 

\subsection{The anomalous dimensions for seesaw type-III}
\label{app:AnoIII}
{\allowdisplaybreaks \begin{align} 
\gamma_{\hat{q}}^{(1)}  = \, &
-\frac{1}{30} \Big(45 g_{2}^{2}  + 80 g_{3}^{2}  + g_{1}^{2}\Big){\bf 1}  + {Y_{d}^{\dagger}  Y_d} + {Y_{u}^{\dagger}  Y_u}\\ 
\gamma_{\hat{q}}^{(2)}  = \, &
+\frac{4}{5} g_{1}^{2} {Y_{u}^{\dagger}  Y_u} -2 {Y_{d}^{\dagger}  Y_d  Y_{d}^{\dagger}  Y_d} -2 {Y_{d}^{\dagger}  Y_{x}^{T}  Y_x^*  Y_d} -2 {Y_{u}^{\dagger}  Y_u  Y_{u}^{\dagger}  Y_u} +{\bf 1} \Big[\frac{199}{900} g_{1}^{4} +\frac{1}{10} g_{1}^{2} g_{2}^{2} +\frac{15}{4} g_{2}^{4} \nonumber \\ 
 & +\frac{8}{45} g_{1}^{2} g_{3}^{2} +8 g_{2}^{2} g_{3}^{2} -\frac{8}{9} g_{3}^{4} +8 g_{3}^{4} N_{G_M} +3 g_{2}^{4} N_{W_M} + \left(\frac{1}{12} g_{1}^{4} +\frac{9}{4} g_{2}^{4} +\frac{8}{3} g_{3}^{4}\right)\tilde{N}_{X_M} \Big] \nonumber \\ 
 &  -\frac{3}{10} {Y_{u}^{\dagger}  Y_u} \,\mbox{Tr}\Big({Y_b  Y_{b}^{\dagger}}\Big) +{Y_{d}^{\dagger}  Y_d} \Big(-3 \,\mbox{Tr}\Big({Y_d  Y_{d}^{\dagger}}\Big)  + \frac{2}{5} g_{1}^{2}  - \,\mbox{Tr}\Big({Y_e  Y_{e}^{\dagger}}\Big) \Big)-3 {Y_{u}^{\dagger}  Y_u} \,\mbox{Tr}\Big({Y_u  Y_{u}^{\dagger}}\Big) \nonumber \\ 
 &-\frac{3}{2} {Y_{u}^{\dagger}  Y_u} \,\mbox{Tr}\Big({Y_w  Y_{w}^{\dagger}}\Big) -3 {Y_{u}^{\dagger}  Y_u} \,\mbox{Tr}\Big({Y_x  Y_{x}^{\dagger}}\Big) \\ 
\gamma_{\hat{l}}^{(1)}  = \, &
\frac{1}{10} \Big(10 {Y_{e}^{\dagger}  Y_e}  + 15 {Y_{w}^{\dagger}  Y_w}  -3 \Big(5 g_{2}^{2}  + g_{1}^{2}\Big){\bf 1}  + 3 {Y_{b}^{\dagger}  Y_b} \Big)\\ 
\gamma_{\hat{l}}^{(2)}  = \, &
\frac{1}{200} \Big[240 g_{1}^{2} {Y_{e}^{\dagger}  Y_e} +1200 g_{2}^{2} {Y_{w}^{\dagger}  Y_w} -36 {Y_{b}^{\dagger}  Y_b  Y_{b}^{\dagger}  Y_b} -60 {Y_{b}^{\dagger}  Y_b  Y_{w}^{\dagger}  Y_w} -400 {Y_{e}^{\dagger}  Y_e  Y_{e}^{\dagger}  Y_e} \nonumber \\ 
 &-45 {Y_{w}^{\dagger}  Y_w  Y_{b}^{\dagger}  Y_b} -300 {Y_{w}^{\dagger}  Y_w  Y_{w}^{\dagger}  Y_w} +6 {\bf 1} \Big(100 g_{2}^{4} N_{W_M}  + 125 g_{2}^{4}  + 25 \Big(3 g_{2}^{4}  + g_{1}^{4}\Big)\tilde{N}_{X_M}  \nonumber \\ 
 &  + 30 g_{1}^{2} g_{2}^{2}  + 69 g_{1}^{4}  \Big)-18 {Y_{b}^{\dagger}  Y_b} \,\mbox{Tr}\Big({Y_b  Y_{b}^{\dagger}}\Big) -90 {Y_{w}^{\dagger}  Y_w} \,\mbox{Tr}\Big({Y_b  Y_{b}^{\dagger}}\Big) -600 {Y_{e}^{\dagger}  Y_e} \,\mbox{Tr}\Big({Y_d  Y_{d}^{\dagger}}\Big) \nonumber \\ 
 &-200 {Y_{e}^{\dagger}  Y_e} \,\mbox{Tr}\Big({Y_e  Y_{e}^{\dagger}}\Big) -180 {Y_{b}^{\dagger}  Y_b} \,\mbox{Tr}\Big({Y_u  Y_{u}^{\dagger}}\Big) -900 {Y_{w}^{\dagger}  Y_w} \,\mbox{Tr}\Big({Y_u  Y_{u}^{\dagger}}\Big)-90 {Y_{b}^{\dagger}  Y_b} \,\mbox{Tr}\Big({Y_w  Y_{w}^{\dagger}}\Big)  \nonumber \\ 
 &-450 {Y_{w}^{\dagger}  Y_w} \,\mbox{Tr}\Big({Y_w  Y_{w}^{\dagger}}\Big) -180 {Y_{b}^{\dagger}  Y_b} \,\mbox{Tr}\Big({Y_x  Y_{x}^{\dagger}}\Big) -900 {Y_{w}^{\dagger}  Y_w} \,\mbox{Tr}\Big({Y_x  Y_{x}^{\dagger}}\Big) \Big]\\ 
\gamma_{\hat{H}_d}^{(1)}  = \, &
3 \,\mbox{Tr}\Big({Y_d  Y_{d}^{\dagger}}\Big)  -\frac{3}{10} \Big(5 g_{2}^{2}  + g_{1}^{2}\Big) + \,\mbox{Tr}\Big({Y_e  Y_{e}^{\dagger}}\Big)\\ 
\gamma_{\hat{H}_d}^{(2)}  = \, &
+\frac{207}{100} g_{1}^{4} +\frac{9}{10} g_{1}^{2} g_{2}^{2} +\frac{15}{4} g_{2}^{4} +3 g_{2}^{4} N_{W_M} +\frac{3}{4} \Big(3 g_{2}^{4}  + g_{1}^{4}\Big)\tilde{N}_{X_M}-9 \,\mbox{Tr}\Big({Y_d  Y_{d}^{\dagger}  Y_d  Y_{d}^{\dagger}}\Big) \nonumber \\ 
 &-\frac{2}{5} g_{1}^{2} \,\mbox{Tr}\Big({Y_d  Y_{d}^{\dagger}}\Big) +16 g_{3}^{2} \,\mbox{Tr}\Big({Y_d  Y_{d}^{\dagger}}\Big) +\frac{6}{5} g_{1}^{2} \,\mbox{Tr}\Big({Y_e  Y_{e}^{\dagger}}\Big) -\frac{3}{10} \,\mbox{Tr}\Big({Y_b  Y_{e}^{\dagger}  Y_e  Y_{b}^{\dagger}}\Big)  \nonumber \\ 
 &-6 \,\mbox{Tr}\Big({Y_d  Y_{d}^{\dagger}  Y_{x}^{T}  Y_x^*}\Big) -3 \,\mbox{Tr}\Big({Y_d  Y_{u}^{\dagger}  Y_u  Y_{d}^{\dagger}}\Big) -3 \,\mbox{Tr}\Big({Y_e  Y_{e}^{\dagger}  Y_e  Y_{e}^{\dagger}}\Big) -\frac{3}{2} \,\mbox{Tr}\Big({Y_e  Y_{w}^{\dagger}  Y_w  Y_{e}^{\dagger}}\Big)   \\ 
\gamma_{\hat{H}_u}^{(1)}  = \, &
-\frac{3}{10} \Big(-10 \,\mbox{Tr}\Big({Y_u  Y_{u}^{\dagger}}\Big)  -10 \,\mbox{Tr}\Big({Y_x  Y_{x}^{\dagger}}\Big)  + 5 g_{2}^{2}  -5 \,\mbox{Tr}\Big({Y_w  Y_{w}^{\dagger}}\Big)  - \,\mbox{Tr}\Big({Y_b  Y_{b}^{\dagger}}\Big)  + g_{1}^{2}\Big)\\ 
\gamma_{\hat{H}_u}^{(2)}  = \, &
+\frac{207}{100} g_{1}^{4} +\frac{9}{10} g_{1}^{2} g_{2}^{2} +\frac{15}{4} g_{2}^{4} +3 g_{2}^{4} N_{W_M} +\frac{3}{4} \Big(3 g_{2}^{4}  + g_{1}^{4}\Big)\tilde{N}_{X_M} +\frac{4}{5} g_{1}^{2} \,\mbox{Tr}\Big({Y_u  Y_{u}^{\dagger}}\Big) \nonumber \\ 
 &+16 g_{3}^{2} \,\mbox{Tr}\Big({Y_u  Y_{u}^{\dagger}}\Big) +6 g_{2}^{2} \,\mbox{Tr}\Big({Y_w  Y_{w}^{\dagger}}\Big) +2 g_{1}^{2} \,\mbox{Tr}\Big({Y_x  Y_{x}^{\dagger}}\Big) +16 g_{3}^{2} \,\mbox{Tr}\Big({Y_x  Y_{x}^{\dagger}}\Big)  \nonumber \\ 
 &-\frac{3}{10} \,\mbox{Tr}\Big({Y_b  Y_{e}^{\dagger}  Y_e  Y_{b}^{\dagger}}\Big) -\frac{57}{40} \,\mbox{Tr}\Big({Y_b  Y_{w}^{\dagger}  Y_w  Y_{b}^{\dagger}}\Big) -6 \,\mbox{Tr}\Big({Y_d  Y_{d}^{\dagger}  Y_{x}^{T}  Y_x^*}\Big) -3 \,\mbox{Tr}\Big({Y_d  Y_{u}^{\dagger}  Y_u  Y_{d}^{\dagger}}\Big) \nonumber \\ 
 &-\frac{3}{2} \,\mbox{Tr}\Big({Y_e  Y_{w}^{\dagger}  Y_w  Y_{e}^{\dagger}}\Big) -9 \,\mbox{Tr}\Big({Y_u  Y_{u}^{\dagger}  Y_u  Y_{u}^{\dagger}}\Big) -\frac{15}{4} \,\mbox{Tr}\Big({Y_w  Y_{w}^{\dagger}  Y_w  Y_{w}^{\dagger}}\Big) -9 \,\mbox{Tr}\Big({Y_x  Y_{x}^{\dagger}  Y_x  Y_{x}^{\dagger}}\Big) \nonumber \\ &  -\frac{27}{100} \,\mbox{Tr}\Big({Y_b  Y_{b}^{\dagger}  Y_b  Y_{b}^{\dagger}}\Big) \\
\gamma_{\hat{d}}^{(1)}  = \, &
2 \Big({Y_{x}^{\dagger}  Y_x} + {Y_d^*  Y_{d}^{T}}\Big) -\frac{2}{15} \Big(20 g_{3}^{2}  + g_{1}^{2}\Big){\bf 1} \\ 
\gamma_{\hat{d}}^{(2)}  = \, &
+\frac{2}{5} g_{1}^{2} {Y_d^*  Y_{d}^{T}} +6 g_{2}^{2} {Y_d^*  Y_{d}^{T}} -2 {Y_{x}^{\dagger}  Y_x  Y_{x}^{\dagger}  Y_x} -2 {Y_d^*  Y_{d}^{T}  Y_d^*  Y_{d}^{T}} -2 {Y_d^*  Y_{u}^{T}  Y_u^*  Y_{d}^{T}} \nonumber \\ 
 &+\frac{1}{225} {\bf 1} \Big[160 g_{1}^{2} g_{3}^{2}  + 1800 g_{3}^{4} N_{G_M}  -200 g_{3}^{4}  + 202 g_{1}^{4}   + 75 \Big(8 g_{3}^{4}  + g_{1}^{4}\Big)\tilde{N}_{X_M}  \Big] \nonumber \\ 
 &-6 {Y_d^*  Y_{d}^{T}} \,\mbox{Tr}\Big({Y_d  Y_{d}^{\dagger}}\Big) -2 {Y_d^*  Y_{d}^{T}} \,\mbox{Tr}\Big({Y_e  Y_{e}^{\dagger}}\Big) +{Y_{x}^{\dagger}  Y_x} \Big(2 g_{1}^{2}  -3 \,\mbox{Tr}\Big({Y_w  Y_{w}^{\dagger}}\Big)  + 6 g_{2}^{2}  \nonumber \\  & -6 \,\mbox{Tr}\Big({Y_u  Y_{u}^{\dagger}}\Big)  -6 \,\mbox{Tr}\Big({Y_x  Y_{x}^{\dagger}}\Big)  -\frac{3}{5} \,\mbox{Tr}\Big({Y_b  Y_{b}^{\dagger}}\Big) \Big)\\ 
\gamma_{\hat{u}}^{(1)}  = \, &
2 {Y_u^*  Y_{u}^{T}}  -\frac{8}{15} \Big(5 g_{3}^{2}  + g_{1}^{2}\Big){\bf 1} \\ 
\gamma_{\hat{u}}^{(2)}  = \, &
-2 \Big({Y_u^*  Y_{d}^{T}  Y_d^*  Y_{u}^{T}} + {Y_u^*  Y_{u}^{T}  Y_u^*  Y_{u}^{T}}\Big)+\frac{4}{225} {\bf 1} \Big[160 g_{1}^{2} g_{3}^{2}  + 214 g_{1}^{4}  + 450 g_{3}^{4} N_{G_M} -50 g_{3}^{4}  \nonumber \\ 
 & + 75 \Big(2 g_{3}^{4}  + g_{1}^{4}\Big)\tilde{N}_{X_M} \Big]-\frac{1}{5} {Y_u^*  Y_{u}^{T}} \Big[15 \,\mbox{Tr}\Big({Y_w  Y_{w}^{\dagger}}\Big)  + 2 g_{1}^{2}  -30 g_{2}^{2}  + 30 \,\mbox{Tr}\Big({Y_u  Y_{u}^{\dagger}}\Big) \nonumber \\ 
 & + 30 \,\mbox{Tr}\Big({Y_x  Y_{x}^{\dagger}}\Big)  + 3 \,\mbox{Tr}\Big({Y_b  Y_{b}^{\dagger}}\Big) \Big]\\ 
\gamma_{\hat{e}}^{(1)}  = \, &
2 {Y_e^*  Y_{e}^{T}}  -\frac{6}{5} g_{1}^{2} {\bf 1} \\ 
\gamma_{\hat{e}}^{(2)}  = \, &
-\frac{3}{5} {Y_e^*  Y_{b}^{T}  Y_b^*  Y_{e}^{T}} -2 {Y_e^*  Y_{e}^{T}  Y_e^*  Y_{e}^{T}} -3 {Y_e^*  Y_{w}^{T}  Y_w^*  Y_{e}^{T}} +\frac{3}{25} g_{1}^{4} {\bf 1} \Big(25 \tilde{N}_{X_M}  + 78\Big)\nonumber \\ 
 &+{Y_e^*  Y_{e}^{T}} \Big[-2 \,\mbox{Tr}\Big({Y_e  Y_{e}^{\dagger}}\Big)  + 6 g_{2}^{2}  -6 \,\mbox{Tr}\Big({Y_d  Y_{d}^{\dagger}}\Big)  -\frac{6}{5} g_{1}^{2} \Big]\\ 
\gamma_{\hat{W}_M}^{(1)}  = \, &
-4 g_{2}^{2} {\bf 1}  + {Y_w^*  Y_{w}^{T}}\\ 
\gamma_{\hat{W}_M}^{(2)}  = \, &
+2 g_{2}^{4} {\bf 1} \Big(3 \tilde{N}_{X_M}  + 4 N_{W_M}  + 10\Big)+\frac{1}{10} \Big[-3 {Y_w^*  Y_{b}^{T}  Y_b^*  Y_{w}^{T}} -10 {Y_w^*  Y_{e}^{T}  Y_e^*  Y_{w}^{T}} \nonumber \\ 
 &-15 {Y_w^*  Y_{w}^{T}  Y_w^*  Y_{w}^{T}} +{Y_w^*  Y_{w}^{T}} \Big\{-10 g_{2}^{2}  -15 \,\mbox{Tr}\Big({Y_w  Y_{w}^{\dagger}}\Big)  -30 \,\mbox{Tr}\Big({Y_u  Y_{u}^{\dagger}}\Big)  -30 \,\mbox{Tr}\Big({Y_x  Y_{x}^{\dagger}}\Big) \nonumber \\ 
 & -3 \,\mbox{Tr}\Big({Y_b  Y_{b}^{\dagger}}\Big)  + 6 g_{1}^{2} \Big\}\Big]\\ 
\gamma_{\hat{G}_M}^{(1)}  = \, &
-6 g_{3}^{2} {\bf 1} \\ 
\gamma_{\hat{G}_M}^{(2)}  = \, &
6 g_{3}^{4} {\bf 1} \Big(3 N_{G_M}  + 3 + \tilde{N}_{X_M} \Big)\\ 
\gamma_{\hat{B}_M}^{(1)}  = \, &
\frac{3}{5} {Y_b^*  Y_{b}^{T}} \\ 
\gamma_{\hat{B}_M}^{(2)}  = \, &
\frac{3}{50} \Big[-3 {Y_b^*  Y_{b}^{T}  Y_b^*  Y_{b}^{T}} -5 \Big(2 {Y_b^*  Y_{e}^{T}  Y_e^*  Y_{b}^{T}}  + 3 {Y_b^*  Y_{w}^{T}  Y_w^*  Y_{b}^{T}} \Big)\nonumber \\ 
 &+3 {Y_b^*  Y_{b}^{T}} \Big\{10 g_{2}^{2}  -10 \,\mbox{Tr}\Big({Y_u  Y_{u}^{\dagger}}\Big)  -10 \,\mbox{Tr}\Big({Y_x  Y_{x}^{\dagger}}\Big)  + 2 g_{1}^{2}  -5 \,\mbox{Tr}\Big({Y_w  Y_{w}^{\dagger}}\Big)  - \,\mbox{Tr}\Big({Y_b  Y_{b}^{\dagger}}\Big) \Big\}\Big]\\ 
\gamma_{\hat{X}_M}^{(1)}  = \, &
-\frac{1}{6} \Big(16 g_{3}^{2}  + 5 g_{1}^{2}  + 9 g_{2}^{2} \Big){\bf 1} \\ 
\gamma_{\hat{X}_M}^{(2)}  = \, &
\frac{1}{36} {\bf 1} \Big[223 g_{1}^{4} +90 g_{1}^{2} g_{2}^{2} +135 g_{2}^{4} +160 g_{1}^{2} g_{3}^{2} +288 (g_{2}^{2} g_{3}^{2}+g_{3}^{4} N_{G_M}) -32 g_{3}^{4}+108 g_{2}^{4} N_{W_M} \nonumber \\ 
 &+\left(75 g_{1}^{4}  +81 g_{2}^{4}  +96 g_{3}^{4} \right)\tilde{N}_{X_M} \Big]\\ 
\gamma_{\hat{\bar{X}}_M}^{(1)}  = \, &
\frac{1}{6} \Big(- \Big(16 g_{3}^{2}  + 5 g_{1}^{2}  + 9 g_{2}^{2} \Big){\bf 1}  + 6 {Y_x^*  Y_{x}^{T}} \Big)\\ 
\gamma_{\hat{\bar{X}}_M}^{(2)}  = \, &
+\frac{1}{36} {\bf 1} \Big[223 g_{1}^{4} +90 g_{1}^{2} g_{2}^{2} +135 g_{2}^{4} +160 g_{1}^{2} g_{3}^{2} +288 (g_{2}^{2} g_{3}^{2}+ g_{3}^{4} N_{G_M}) -32 g_{3}^{4} +108 g_{2}^{4} N_{W_M} \nonumber \\ 
 &+\left(75 g_{1}^{4} +81 g_{2}^{4} +96 g_{3}^{4}\right)\tilde{N}_{X_M} \Big]+\frac{1}{10} \Big[-20 \Big({Y_x^*  Y_d  Y_{d}^{\dagger}  Y_{x}^{T}} + {Y_x^*  Y_{x}^{T}  Y_x^*  Y_{x}^{T}}\Big)\nonumber \\ 
 &- {Y_x^*  Y_{x}^{T}} \Big\{15 \,\mbox{Tr}\Big({Y_w  Y_{w}^{\dagger}}\Big)  + 30 \,\mbox{Tr}\Big({Y_u  Y_{u}^{\dagger}}\Big)+ 30 \,\mbox{Tr}\Big({Y_x  Y_{x}^{\dagger}}\Big)  + 3 \,\mbox{Tr}\Big({Y_b  Y_{b}^{\dagger}}\Big)  + 4 g_{1}^{2} \Big\}\Big]
\end{align} }

\subsection{Beta coefficients for the seesaw type-III at 2-loop level}
\label{app:BetaIII}
{\allowdisplaybreaks  
\begin{align} 
\beta_{g_1}^{(1)}  = \, &
\frac{1}{10} g_{1}^{3} \Big(25 \tilde{N}_{X_M}  + 66\Big)\\ 
\beta_{g_1}^{(2)}  = \, &
\frac{1}{150} g_{1}^{3} \Big[125 \Big(16 g_{3}^{2}  + 5 g_{1}^{2}  + 9 g_{2}^{2} \Big)\tilde{N}_{X_M} +6 \Big\{199 g_{1}^{2} +135 g_{2}^{2} +440 g_{3}^{2} -9 \,\mbox{Tr}\Big({Y_b  Y_{b}^{\dagger}}\Big) \nonumber \\ 
 &-70 \,\mbox{Tr}\Big({Y_d  Y_{d}^{\dagger}}\Big) -90 \,\mbox{Tr}\Big({Y_e  Y_{e}^{\dagger}}\Big)-130 \,\mbox{Tr}\Big({Y_u  Y_{u}^{\dagger}}\Big) -60 \,\mbox{Tr}\Big({Y_w  Y_{w}^{\dagger}}\Big) -190 \,\mbox{Tr}\Big({Y_x  Y_{x}^{\dagger}}\Big) \Big\}\Big]\\ 
\beta_{g_2}^{(1)}  = \, &
\frac{1}{2} g_{2}^{3} \Big(3 \tilde{N}_{X_M} + 4 N_{W_M}  + 2\Big)\\ 
\beta_{g_2}^{(2)}  = \, &
\frac{1}{30} g_{2}^{3} \Big[54 g_{1}^{2} +750 g_{2}^{2} +720(g_{3}^{2} +g_{2}^{2} N_{W_M}) +15 \Big(16 g_{3}^{2}  + 21 g_{2}^{2}  + 5 g_{1}^{2} \Big)\tilde{N}_{X_M} \nonumber \\ 
 &-18 \,\mbox{Tr}\Big({Y_b  Y_{b}^{\dagger}}\Big)-280 \,\mbox{Tr}\Big({Y_w  Y_{w}^{\dagger}}\Big)  -60 \,\mbox{Tr}\Big({Y_e  Y_{e}^{\dagger}}\Big) -180 \Big(\mbox{Tr}\Big({Y_d  Y_{d}^{\dagger}}\Big) +\mbox{Tr}\Big({Y_u  Y_{u}^{\dagger}}\Big) \nonumber \\
& +\mbox{Tr}\Big({Y_x  Y_{x}^{\dagger}}\Big)\Big) \Big]\\ 
\beta_{g_3}^{(1)}  = \, &
g_{3}^{3} \Big(3 N_{G_M}  -3 + \tilde{N}_{X_M}\Big)\\ 
\beta_{g_3}^{(2)}  = \, &
\frac{1}{15} g_{3}^{3} \Big[33 g_{1}^{2} +135 g_{2}^{2} +210 g_{3}^{2} +810 g_{3}^{2} N_{G_M} +5 \Big(34 g_{3}^{2}  + 5 g_{1}^{2}  + 9 g_{2}^{2} \Big)\tilde{N}_{X_M} \nonumber \\ 
 &+170 g_{3}^{2} N_{\bar{X}_M} -60 \,\mbox{Tr}\Big({Y_d  Y_{d}^{\dagger}}\Big) -60 \,\mbox{Tr}\Big({Y_u  Y_{u}^{\dagger}}\Big) -60 \,\mbox{Tr}\Big({Y_x  Y_{x}^{\dagger}}\Big) \Big]
\end{align}}

\end{document}